\documentclass[superscriptaddress, amsmath, nofootinbib, onecolumn, notitlepage]{revtex4-1}

\usepackage{microtype} 
\usepackage{graphicx}
\usepackage{xcolor}
\usepackage{orcidlink}
\usepackage{enumerate}
\usepackage{enumitem}
\usepackage{ulem}
\definecolor{linkcolor}{rgb}{0.0,0.3,0.5}
\definecolor{linkcolor}{rgb}{0.0,0.3,0.5}
\definecolor{mypurple}{RGB}{143, 116, 210}

\graphicspath{{Figures/}}

\newcommand{\hias}{School of Fundamental Physics and Mathematical Sciences, Hangzhou Institute for Advanced Study, University of Chinese Academy of Sciences, Hangzhou 310024, China}

\newcommand{\kcl}{Theoretical Particle Physics and Cosmology Group,
Physics Department, King's College London, Strand, London WC2R 2LS,
United Kingdom}

\begin{document}

\title{Massive boson stars: Stability and GW emission in head-on mergers}



\author{Bo-Xuan Ge
\orcidlink{0000-0003-0738-3473}}
\email{bo-xuan.ge@ucas.ac.cn}
\affiliation{\hias}
\affiliation{\kcl}

\begin{abstract}
We investigate quartically self--interacting massive boson stars by constructing equilibrium sequences and performing dynamical evolutions. The mass curve $M(|\phi_c|)$ along these sequences develops multiple extrema, yet stability changes only at the first maximum; configurations beyond it become highly compact and collapse under numerically induced perturbations, with near--critical models displaying a short--lived double--dive behaviour. Head--on collisions of equal--mass stars yield three distinct outcomes --- boson star remnants, black hole formation at contact, and collapse of each star to a black hole prior to contact. The associated gravitational-wave energies reflect a competition between increasing compactness, which enhances the efficiency of gravitational-wave emission, and decreasing tidal deformability, which suppresses merger asymmetries, and at large self--interaction strengths the collapse-before-contact branch exhibits a pronounced non--monotonic structure. The simulations reported here constitute a substantial catalogue of initial conditions and waveforms, providing a natural basis for constructing surrogate models capable of rapidly predicting gravitational-wave signals across an extended parameter space.
\end{abstract}

\maketitle

\section{Introduction}
\label{sec:intro}

The first detections of gravitational waves from binary black holes~\cite{Abbott:2016blz,LIGOScientific:2018mvr,LIGOScientific:2020ibl,LIGOScientific:2021djp} have opened a new observational window on strong-field gravity, allowing direct tests of compact-object dynamics and binary formation scenarios~\cite{Abbott:2016blz,LIGOScientific:2018jsj,LIGOScientific:2020kqk,KAGRA:2021duu}.

Gravitational waves also offer a promising avenue to probe the nature of dark matter. In light of the limited progress in direct searches for weakly interacting massive particles (WIMPs), boson stars—with low field masses ($m \lesssim \mathrm{eV}$) and bosonic self-gravity—have emerged as compelling dark-matter candidates~\cite{Guver:2012ba, Alcubierre:2001ea, Hu:2000ke, Macedo:2013qea}. Independently of their role as dark matter, boson stars are widely investigated as horizonless compact objects capable of mimicking several observational signatures commonly associated with black holes and neutron stars~\cite{Sennett:2017etc, Herdeiro:2021lwl, Rosa:2022tfv, Brihaye:2020klz, Rosa:2023qcv, Rosa:2022toh}.

Within general relativity, boson stars have been extensively explored in a variety of contexts, including their 
isolated spacetime structure~\cite{Colpi:1986ye, Seidel:1990jh, Kobayashi:1994qi, Ryan:1996nk, Schunck:1996he, Balakrishna:1997ej, Yoshida:1997qf, Schunck:1999zu, Schunck:2003kk, Balakrishna:2006ru, Balakrishna:2007mr, Hartmann:2012da, Siemonsen:2020hcg, Evstafyeva:2025mvx, Marks:2025jpt, Evstafyeva:2023kfg, Marks:2025xxv, Marks:2025jit, Ma:2024olw, Ding:2023syj, Liang:2022mjo, Zhang:2023qxf, Zhang:2025xnl, deSa:2025nsx, Herdeiro:2025lwf, Herdeiro:2024myz, Ildefonso:2023qty, Brito:2023fwr}, 
formation mechanisms~\cite{Schunck:1999pm, Sanchis-Gual:2019ljs, Siemonsen:2023hko}, 
and binary dynamics~\cite{Palenzuela:2006wp, Palenzuela:2007dm, Palenzuela:2017kcg, Helfer:2021brt, Sanchis-Gual:2020mzb, Bezares:2022obu, Cardoso:2022vpj, Croft:2022bxq, Sanchis-Gual:2022zsr, Evstafyeva:2022bpr, Siemonsen:2023age, Ge:2024fum, Evstafyeva:2024qvp, Brito:2025rld, Jaramillo:2022zwg, Damour:2025oys}.

In particular, our recent work~\cite{Ge:2024itl} carried out a two-parameter survey of head-on collisions of equal-mass mini and solitonic boson star binaries. The resulting gravitational-wave emission exhibits highly structured behaviour: the radiated energies can exceed those from comparable black-hole collisions by an order of magnitude, display needle-sharp features, and show abrupt jumps to black-hole levels. These patterns reflect how the scalar-field potential shapes the structure of the corresponding families of static boson star solutions. The choice of potential determines the shape of the mass curve, the variation of compactness with central amplitude, the degree of deformability, and the location of dynamically stable and unstable segments along the sequence. These properties in turn influence the merger dynamics and the gravitational-wave output: higher compactness tends to enhance the efficiency of gravitational-wave emission, whereas greater deformability can increase the merger asymmetry and thereby also enhance the emitted radiation. The resulting behaviour is therefore non-trivial, since these effects do not vary monotonically or in tandem along the sequence.

Despite this progress, the dynamics of massive boson stars with quartic self-interactions remain comparatively unexplored. Such models possess qualitatively different internal structure and stability properties, and their head-on collisions may imprint distinct gravitational-wave signatures that have not yet been systematically analysed. The aim of this work is to address this gap.

In this paper we extend the framework of Ref.~\cite{Ge:2024itl} to quartically self-interacting massive boson stars, and investigate both their single-star stability and the gravitational-wave emission from head-on collisions.

The paper is organized as follows. In Sec.~\ref{sec:BSmodel} we introduce the massive boson star model and the relevant field equations. Section~\ref{sec:2DBScode} summarizes the Cartoon-based numerical framework used in our evolutions. In Sec.~\ref{sec:SBS} we analyse families of static massive boson star solutions and their dynamical stability, and in Sec.~\ref{sec:BSC} we investigate head-on binary collisions and their gravitational-wave emission. Our main findings are summarized in Sec.~\ref{sec:summary}, followed by an outlook on future AI-assisted developments in Sec.~\ref{sec:outlook}. Throughout this work, we employ geometric units where $G = c = 1$.

\section{Massive Boson Star Model}
\label{sec:BSmodel}

The boson stars under consideration are described by a complex Klein--Gordon scalar field,
\begin{equation}
  \phi(t, r) = |\phi(r)| e^{\mathrm{i} \omega t},
  \label{eq:phidecomp}
\end{equation}
where $|\phi(r)|$ is the scalar-field amplitude and $\omega$ the frequency. The field is minimally coupled to gravity. The associated action combines the Einstein--Hilbert term and the matter contribution in curved spacetime,
\begin{equation}
  S = \int_{M} \left[ \mathcal{L}_{EH} + \mathcal{L}_M \right] \sqrt{-g} \, d^4x,
\end{equation}
where 
\begin{equation}
\mathcal{L}_{EH} = \frac{R}{16 \pi}
\end{equation}
represents the Einstein--Hilbert Lagrangian density, and 
\begin{equation}
\mathcal{L}_M = -\frac{1}{2} g^{\mu \nu} \nabla_\mu \bar{\phi} \nabla_\nu \phi - \frac{1}{2} V(|\phi|^2)
\end{equation}
is the Lagrangian density of the scalar field. Here, \(\bar{\phi}\) denotes the complex conjugate of the scalar field $\phi$, and \(V(|\phi|^2)\) is a self-interaction potential that depends only on the modulus of the field, reflecting the global \(U(1)\) symmetry in the complex plane. 

Different choices of potential lead to different boson star models. The simplest example is the mini-boson star, which contains only a mass term and no self-interaction; this case has already been discussed in detail in Ref.~\cite{Ge:2024itl}. In this work we instead focus on massive boson stars~\cite{PhysRevLett.57.2485} with a quartic self-interaction potential,
\begin{equation}\label{massiveV}
V=m^2|\phi|^2 + \frac{\lambda}{2}|\phi|^4
,\end{equation}
where $\lambda$ is a dimensionless coupling constant.

Varying the action with respect to the metric and the scalar field yields, respectively, the Einstein field equations and the Klein--Gordon equation in curved spacetime,
\begin{equation}\label{3.2}
R_{\mu \nu}-\frac{1}{2} R g_{\mu \nu}=8 \pi T_{\mu \nu}
,\end{equation}
\begin{equation}\label{3.3}
g^{\mu \nu} \nabla_\mu \nabla_\nu \phi=\frac{\partial V}{\partial|\phi|^2} \phi
,\end{equation}
where the stress--energy tensor of the scalar field is
\begin{equation}
T_{\mu \nu}=\frac{1}{2} \nabla_\mu \bar{\phi} \nabla_\nu \phi+\frac{1}{2} \nabla_\nu \bar{\phi} \nabla_\mu \phi-\frac{1}{2} g_{\mu \nu}\left[g^{\alpha \beta} \nabla_\alpha \bar{\phi} \nabla_\beta \phi+V\right].
\end{equation}
A detailed account of the construction of these boson star solutions can be found in Ref.~\cite{Helfer:2021brt}.

\section{2-Dimensional Boson Stars Code}
\label{sec:2DBScode}
\begin{figure}
\centering
\includegraphics[width=0.4\textwidth]{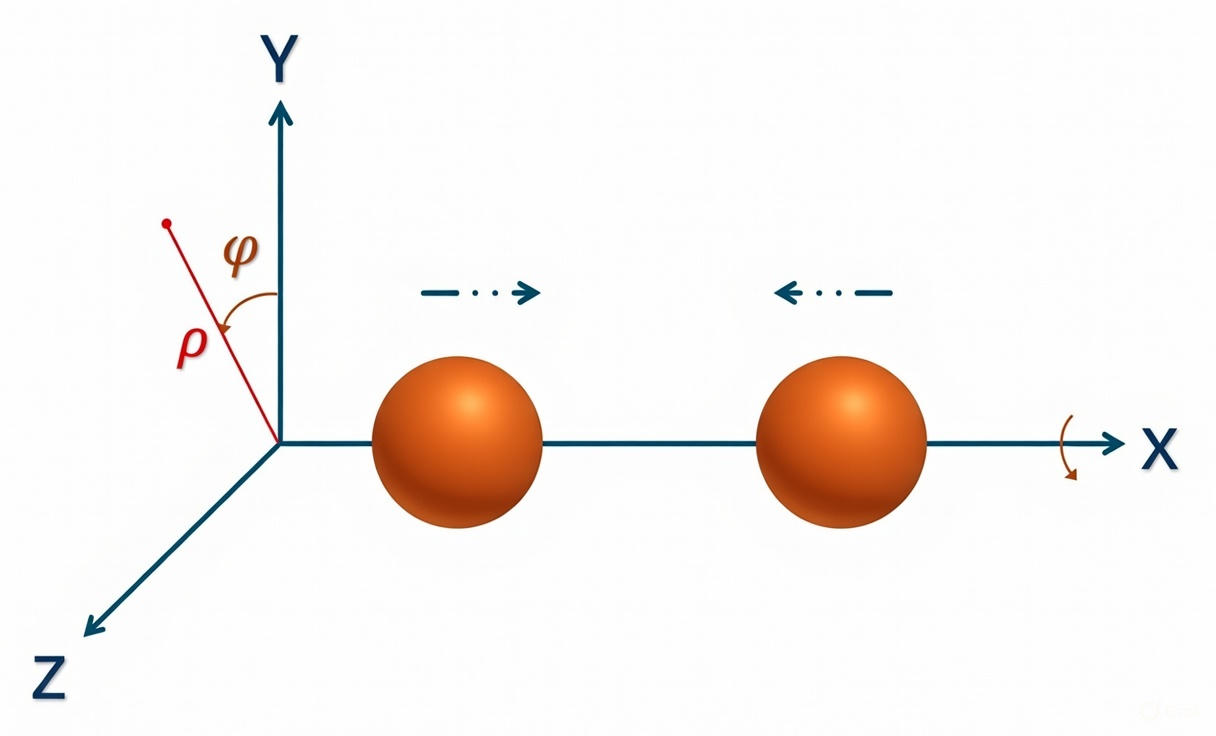}
\caption{Illustration of the evolution scheme used in the two-dimensional boson star code. 
The evolution is performed in Cartesian coordinates \((x,y,z)\) under the assumption of rotational symmetry about the \(x\)-axis. 
The Cartoon reduction reconstructs the full three-dimensional solution from data on the two-dimensional \((x,\rho)\) plane, represented here by the \(z=0\) plane with \(y=\rho\ge0\). 
The positive-\(x\) interval shown in the figure reflects the chosen placement of the finite computational domain along the symmetry axis.}\label{Cartoon3}
\end{figure}

The numerical simulations presented in this work are performed with the numerical relativity code {\sc GRChombo}~\cite{Clough:2015sqa, Radia:2021smk, Andrade:2021rbd}, which employs adaptive mesh refinement (AMR). Although both the coordinate basis and the grid layout are defined in Cartesian coordinates, spacetimes with symmetries can still be simulated efficiently by imposing appropriate symmetry conditions. The \textbf{\textit{Cartoon method}} is designed precisely around this idea. Initially proposed by Alcubierre et al.~\cite{Alcubierre:1999ab} for vacuum spacetimes, it was later extended by Shibata to systems coupled to hydrodynamics, where it proved both stable and accurate in simulations of rotating stellar collapse~\cite{Shibata:2000gt}. More recently, the Cartoon method has been generalized to studies in higher-dimensional spacetimes~\cite{Shibata:2010wz, Cook:2016soy}.

In this work, we employ the Cartoon method to perform an effective dimensional reduction from three to two dimensions, thereby significantly improving computational efficiency. Our implementation follows the efficient framework proposed by Cook et al.~\cite{Cook:2016soy, Cook:2016qnt}, rather than the original formulation of Alcubierre et al.~\cite{Alcubierre:1999ab}.

We use \((x,y,z)\) to denote Cartesian coordinates and assume rotational symmetry about the \(x\)-axis. We then introduce cylindrical coordinates \((\rho,\varphi,x)\), as shown in Fig.~1. By evolving the data on the plane \(z=0\), with \(y=\rho\ge0\), the complete three-dimensional solution is reconstructed by rotation in the \(\varphi\) direction.

The use of a positive \(x\)-interval is unrelated to the Cartoon reduction. It is a choice of coordinate origin and computational-domain placement along the symmetry axis. In the binary simulations, the two stars are placed at \(x_{\rm L}=x_{\rm c}-D/2\) and \(x_{\rm R}=x_{\rm c}+D/2\), with \(x_{\rm c}\) chosen so that the whole binary lies inside the positive-\(x\) part of the grid.

The transformation between the two coordinate systems is
\begin{equation}
{\setlength{\jot}{6pt}%
\begin{aligned}
y&=\rho \cos \varphi,   \\
z&=\rho \sin \varphi,  \\
\rho&=\sqrt{y^2+z^2}, \\
\varphi&=\arctan \frac{z}{y}.
\end{aligned}}
\end{equation}
We use the following notation for tensor indices:
\begin{equation}
{\setlength{\jot}{6pt}%
\begin{aligned}
A,B,C,\ldots &= t,x,y,z, \\
\bar A,\bar B,\bar C,\ldots &= t,x,\rho,\varphi, \\
I,J,K,\ldots &= x,y,z, \\
i,j,k,\ldots &= x,y .
\end{aligned}}
\end{equation}
In cylindrical coordinates, the axial Killing field associated with rotational symmetry about the $x$-axis is $\xi=\partial_\varphi$. Axisymmetry implies $\mathcal{L}_\xi g = 0$, which, in the cylindrical (barred) coordinate basis, is equivalent to $\partial_\varphi g_{\bar A \bar B} = 0$.

We now apply the same symmetry argument to a \textbf{symmetric} rank-$2$ tensor $T_{AB}$. 
For our \textbf{non-rotating} axisymmetric configurations, $T_{x\varphi}=0$.
Using the tensor transformation law, we find
\begin{equation}\label{5.9}
{\setlength{\jot}{6pt}%
\begin{aligned}
T_{x \varphi} & =\frac{\partial X^A}{\partial x} \frac{\partial X^B}{\partial \varphi} T_{A B} =-z T_{x y}+y T_{x z}
.\end{aligned}}
\end{equation}
Thus,
$T_{x z}=\frac{z}{y} T_{x y}.$
Using a similar argument, we obtain
\begin{equation}
\begin{aligned}
T_{\rho \varphi} & =\frac{\partial X^A}{\partial \rho} \frac{\partial X^B}{\partial \varphi} T_{A B} \\ & =-z \cos \varphi T_{y y}+y \cos \varphi T_{y z}-z \sin \varphi T_{z y}+y \sin \varphi T_{z z}
.\end{aligned}
\end{equation}
This yields
\begin{equation}\label{5.11}
{\setlength{\jot}{6pt}%
\begin{aligned}
\left(\frac{y^2}{\rho}-\frac{z^2}{\rho}\right) T_{y z} & =\frac{y z}{\rho}\left(T_{y y}-T_{z z}\right), \\ T_{y z} & =\frac{y z}{y^2 - z^2}\left(T_{y y}-T_{z z}\right) .
\end{aligned}}
\end{equation}
Since we work on the plane $z=0$, Eq.~\eqref{5.9} gives $T_{xz}=0$ and Eq.~\eqref{5.11} gives $T_{yz}=0$, hence $T_{iz}=0$.

The same axial Killing field can also be written as
\begin{equation}
\xi^A = y\left(\frac{\partial}{\partial z}\right)^A - z\left(\frac{\partial}{\partial y}\right)^A
.\end{equation}
Within the \( z = 0 \) hyperplane, its Lie derivative acting on \( T_{i z} \) is
\begin{equation}
{\setlength{\jot}{6pt}%
\begin{aligned}
\mathcal{L}_{\xi} T_{i z} & =\xi^A \partial_A T_{i z}+T_{A z} \partial_i \xi^A+T_{i A} \partial_z \xi^A \\
&= y \partial_z T_{i z}+T_{z z} \delta_{i y}-T_{i y}.
\end{aligned}}
\end{equation}

Imposing axisymmetry, \(\mathcal{L}_{\boldsymbol{\xi}} T_{i z} = 0\), then gives
\begin{equation}
\partial_z T_{i z}=\frac{T_{i y}-T_{z z} \delta_{i y}}{y}
.\end{equation}
The same procedure can be straightforwardly extended to other tensorial quantities and their corresponding derivatives. Detailed derivations and a complete set of relations can be found in Ref.~\cite{Cook:2016soy}. Here we present only the modified expressions relevant to our study.
\begin{equation}
{\setlength{\jot}{6pt}%
\begin{aligned}
\partial_z \phi & =\partial_i \partial_z \phi=0, \\
\partial_z^2 \phi & =\frac{\partial_y \phi}{y}, \\
V^z & =\partial_i V^z=\partial_z V^i=\partial_z^2 V^z=0, \\
\partial_z V^z & =\frac{V^y}{y}, \\
\partial_i \partial_z V^z & =\left(\frac{\partial_i V^y}{y}-\delta_{i y} \frac{V^y}{y^2}\right), \\
\partial_z^2 V^i & =\left(\frac{\partial_y V^i}{y}-\delta_y^i \frac{V^y}{y^2}\right), \\
T_{i z},&~ \partial_z T_{z z},~ \partial_i \partial_z T_{z z},~ \partial_z^2 T_{i z},~ \partial_z T_{i j},~ \partial_i \partial_z T_{j k}=0.
\end{aligned}}
\end{equation}
Using these relations, we adapt the BSSN formalism and the matter evolution equations to our two-dimensional setup. The BSSN sector is discussed in detail in Ref.~\cite{Cook:2016soy}. Below we quote only the expressions for the matter fields and their evolution equations that are relevant for this work; the detailed derivations are provided in Appendix~\ref{app:2dbsequ}. For the complex scalar field, we write
\[
\phi = \phi_{\mathrm{Re}} + i\phi_{\mathrm{Im}}, \qquad
\Pi = \Pi_{\mathrm{Re}} + i\Pi_{\mathrm{Im}},
\]
where $\Pi$ denotes the standard first-order-in-time variable introduced below.

\begin{equation}\label{eq:2dbsequ}
{\setlength{\jot}{9pt}%
\begin{aligned}
\partial_t \phi_{\mathrm{Re}}
&= \beta^i \partial_i \phi_{\mathrm{Re}} - \alpha\, \Pi_{\mathrm{Re}} ,\\
\partial_t \phi_{\mathrm{Im}}
&= \beta^i \partial_i \phi_{\mathrm{Im}} - \alpha\, \Pi_{\mathrm{Im}} ,\\
\partial_t \Pi_{\mathrm{Re}} 
&= \beta^i \partial_i \Pi_{\mathrm{Re}}
  - \chi \tilde{\gamma}^{ij}\partial_i\alpha\,\partial_j\phi_{\mathrm{Re}}
  + \alpha\Big[
      K \Pi_{\mathrm{Re}}
      + V^{\prime} \phi_{\mathrm{Re}}
      + \frac{1}{2}\tilde{\gamma}^{ij}\partial_i\chi\,\partial_j\phi_{\mathrm{Re}}
      + \chi\Big(
          \Gamma^i \partial_i \phi_{\mathrm{Re}}
          - \tilde{\gamma}^{ij}\partial_i\partial_j\phi_{\mathrm{Re}}
          - \tilde{\gamma}^{zz}\frac{\partial_y\phi_{\mathrm{Re}}}{y}
        \Big)
    \Big] ,\\
\partial_t \Pi_{\mathrm{Im}} 
&= \beta^i \partial_i \Pi_{\mathrm{Im}}
  - \chi \tilde{\gamma}^{ij}\partial_i\alpha\,\partial_j\phi_{\mathrm{Im}}
  + \alpha\Big[
      K \Pi_{\mathrm{Im}}
      + V^{\prime} \phi_{\mathrm{Im}}
      + \frac{1}{2}\tilde{\gamma}^{ij}\partial_i\chi\,\partial_j\phi_{\mathrm{Im}}
      + \chi\Big(
          \Gamma^i \partial_i \phi_{\mathrm{Im}}
          - \tilde{\gamma}^{ij}\partial_i\partial_j\phi_{\mathrm{Im}}
          - \tilde{\gamma}^{zz}\frac{\partial_y\phi_{\mathrm{Im}}}{y}
        \Big)
    \Big] ,\\
\rho
&= \frac12\Bigl(
    \Pi_{\mathrm{Re}}^{2} + \Pi_{\mathrm{Im}}^{2}
  + \chi\, \tilde{\gamma}^{ij}
    \bigl(
        \partial_{i}\phi_{\mathrm{Re}}\,\partial_{j}\phi_{\mathrm{Re}}
      + \partial_{i}\phi_{\mathrm{Im}}\,\partial_{j}\phi_{\mathrm{Im}}
    \bigr)
  + V(|\phi|^{2})
  \Bigr) ,\\
S_i
&=\Pi_{\mathrm{Re}}\, \partial_i \phi_{\mathrm{Re}} + \Pi_{\mathrm{Im}}\, \partial_i \phi_{\mathrm{Im}} ,\\
S_{i j}
&= \partial_i \phi_{\mathrm{Re}} \partial_j \phi_{\mathrm{Re}}
+ \partial_i \phi_{\mathrm{Im}} \partial_j \phi_{\mathrm{Im}}
- \frac{1}{2\chi}\,\tilde{\gamma}_{i j}\bigl(|\nabla\phi|^{2} - |\Pi|^{2} + V(|\phi|^{2})\bigr)  ,\\
S_{z z}
&= - \frac{1}{2\chi}\,\tilde{\gamma}_{z z}\bigl(|\nabla\phi|^{2} - |\Pi|^{2} + V(|\phi|^{2})\bigr)  ,\\
S 
&= \chi\,\tilde{\gamma}^{ij}\bigl(\partial_i \phi_{\mathrm{Re}} \partial_j \phi_{\mathrm{Re}} + \partial_i \phi_{\mathrm{Im}} \partial_j \phi_{\mathrm{Im}}\bigr) + \frac{3}{2}\bigl(|\Pi|^{2} - |\nabla\phi|^{2} - V(|\phi|^{2})\bigr) ,
\end{aligned}}
\end{equation}
where
\[
\Pi \equiv -\frac{1}{\alpha}
\left(
\partial_t \phi - \beta^i \partial_i \phi
\right),
\]
$V^{\prime}\equiv\frac{\partial V}{\partial |\phi|^{2}}$,
$\alpha$ is the lapse function,
$\beta^i$ are the components of the shift vector,
$\chi$ is the conformal factor, and quantities marked with a $\sim$ denote conformally transformed quantities.

The corresponding gravitational-wave extraction within the Cartoon method, including the computation of the radiated energy from the extracted signal, is described in detail in Ref.~\cite{Cook:2016qnt}.
%

\section{Single Boson Stars}
\label{sec:SBS}

Before turning to binary configurations, we first discuss the equilibrium sequences—that is, one-parameter families of static boson star solutions labelled by the central scalar amplitude $|\phi_c|$—and the dynamical (in)stability of isolated massive boson stars in the model of Sec.~\ref{sec:BSmodel}. For such one-parameter families of self-gravitating configurations, it is common to adopt a turning–point picture: extrema of the boson star mass $M(|\phi_c|)$ are then expected to be associated with changes in the number of unstable modes.
In this spirit, the condition
\begin{equation}
  \frac{dM}{d|\phi_c|} = 0
\end{equation}
has frequently been used in the boson star literature as a practical stability indicator, in analogy with neutron and white dwarf sequences~\cite{Lee:1988av,Gleiser:1988ih,Liebling:2012fv}.

Recent work has shown, however, that this criterion must be refined for boson stars. A detailed radial–mode analysis of several spherical bosonic models—including mini boson stars, solitonic boson stars, axion stars and Proca stars—demonstrates that the naive statement “stability changes whenever $dM/d|\phi_c| = 0$” is not generally correct~\cite{Santos:2024vdm}. Transitions between radially stable and unstable configurations do occur only at critical points of the sequence, i.e. at extrema of $M(|\phi_c|)$, but not every such critical point corresponds to a zero–frequency radial mode. Beyond the first mass extremum, additional extrema may lie entirely within a radially unstable portion of the sequence and therefore do not separate branches with different numbers of unstable modes~\cite{Santos:2024vdm}.

In this section we show that massive boson stars with quartic self--interaction follow the same qualitative pattern. For fixed self--interaction strength $\lambda$, we construct equilibrium sequences by varying the central scalar amplitude $|\phi_c|$ and computing the corresponding mass $M(|\phi_c|)$. We then evolve these single-star configurations within the numerical-relativity framework using the Cartoon method. No explicit perturbation is added to the initial data; instead, small deviations from exact equilibrium arise from truncation error and residual constraint violations, and these deviations serve as generic perturbations. By monitoring whether a configuration remains long--lived and close to its initial equilibrium profile or instead undergoes collapse or large--amplitude migration, we identify stable and unstable segments along the sequence.

Our results confirm that stability changes occur only at extrema of $M(|\phi_c|)$, and that, for the quartic model considered here, only the \emph{first} extremum corresponds to a change of stability; beyond this point, all subsequent critical points lie within the dynamically unstable regime.
Massive boson stars thus provide an additional, independent example of the universal stability structure identified in Ref.~\cite{Santos:2024vdm}, now extended to the quartically self–interacting scalar model considered here.

\begin{figure*}
  \centering
  \begin{minipage}{0.49\textwidth}
    \centering
    \includegraphics[width=\linewidth]{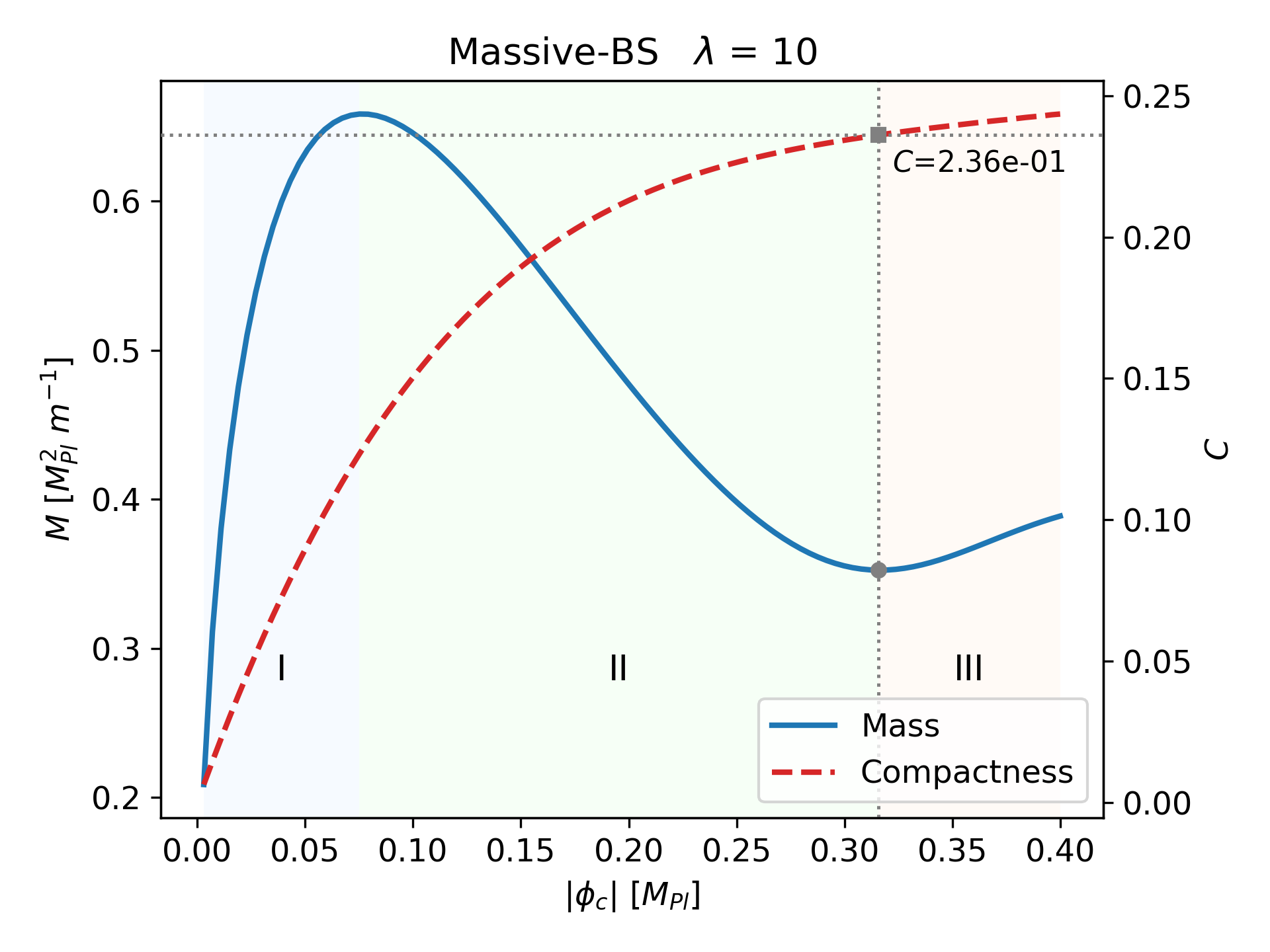}
  \end{minipage}
  \hfill
  \begin{minipage}{0.49\textwidth}
    \centering
    \includegraphics[width=\linewidth]{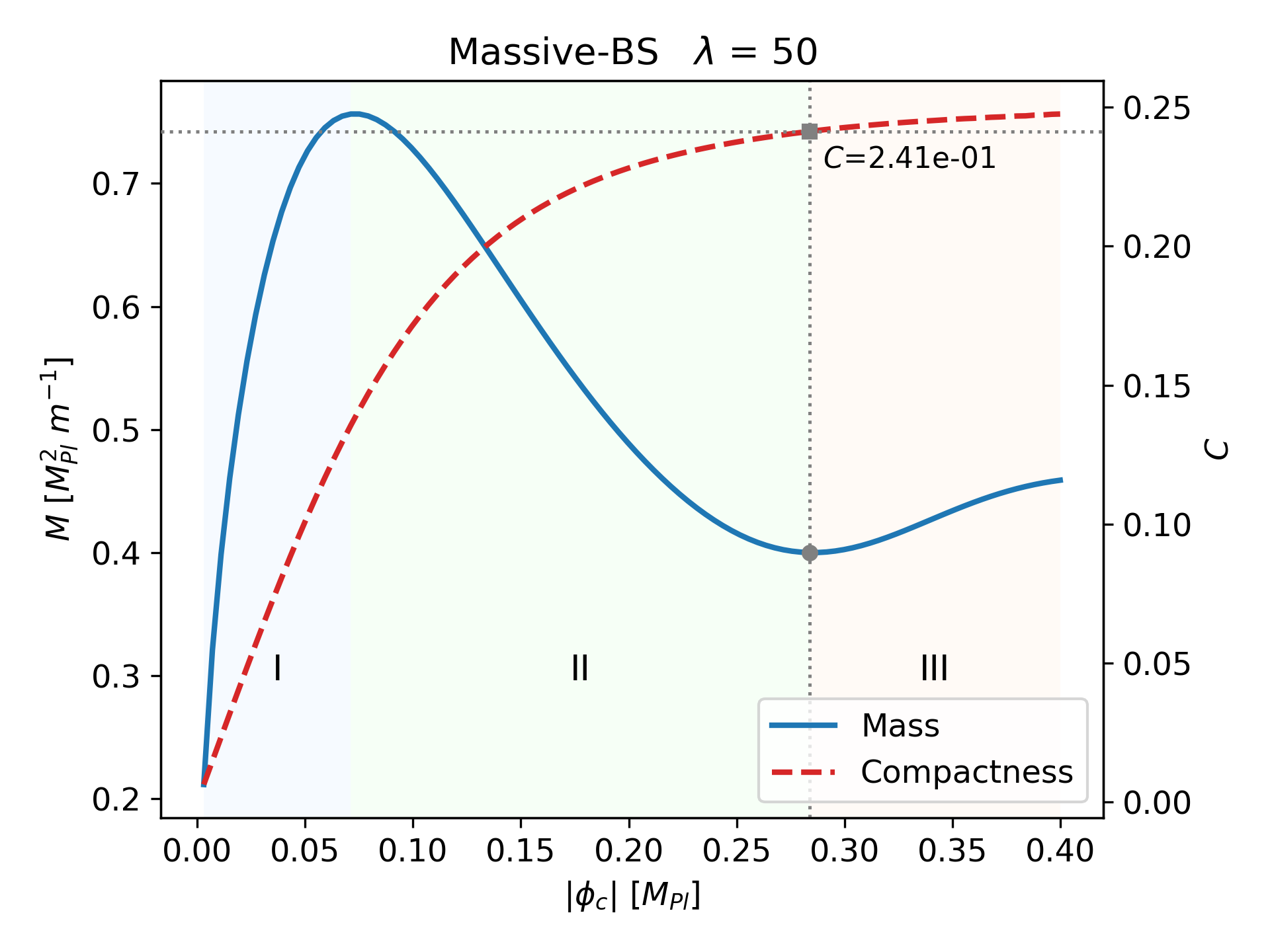}
  \end{minipage}

  \vspace{0.3cm} 

  \begin{minipage}{0.49\textwidth}
    \centering
    \includegraphics[width=\linewidth]{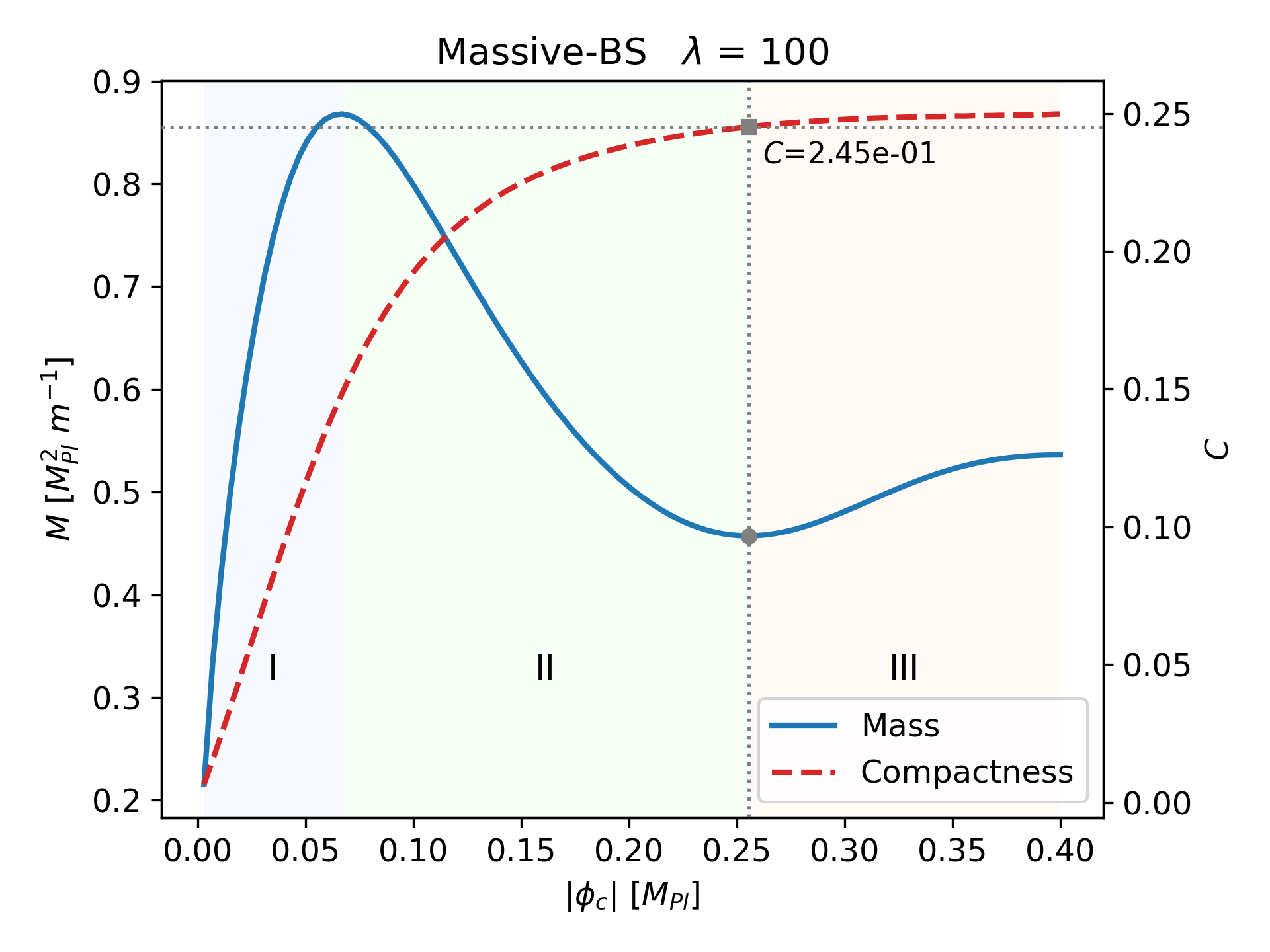}
  \end{minipage}
  \hfill
  \begin{minipage}{0.49\textwidth}
    \centering
    \includegraphics[width=\linewidth]{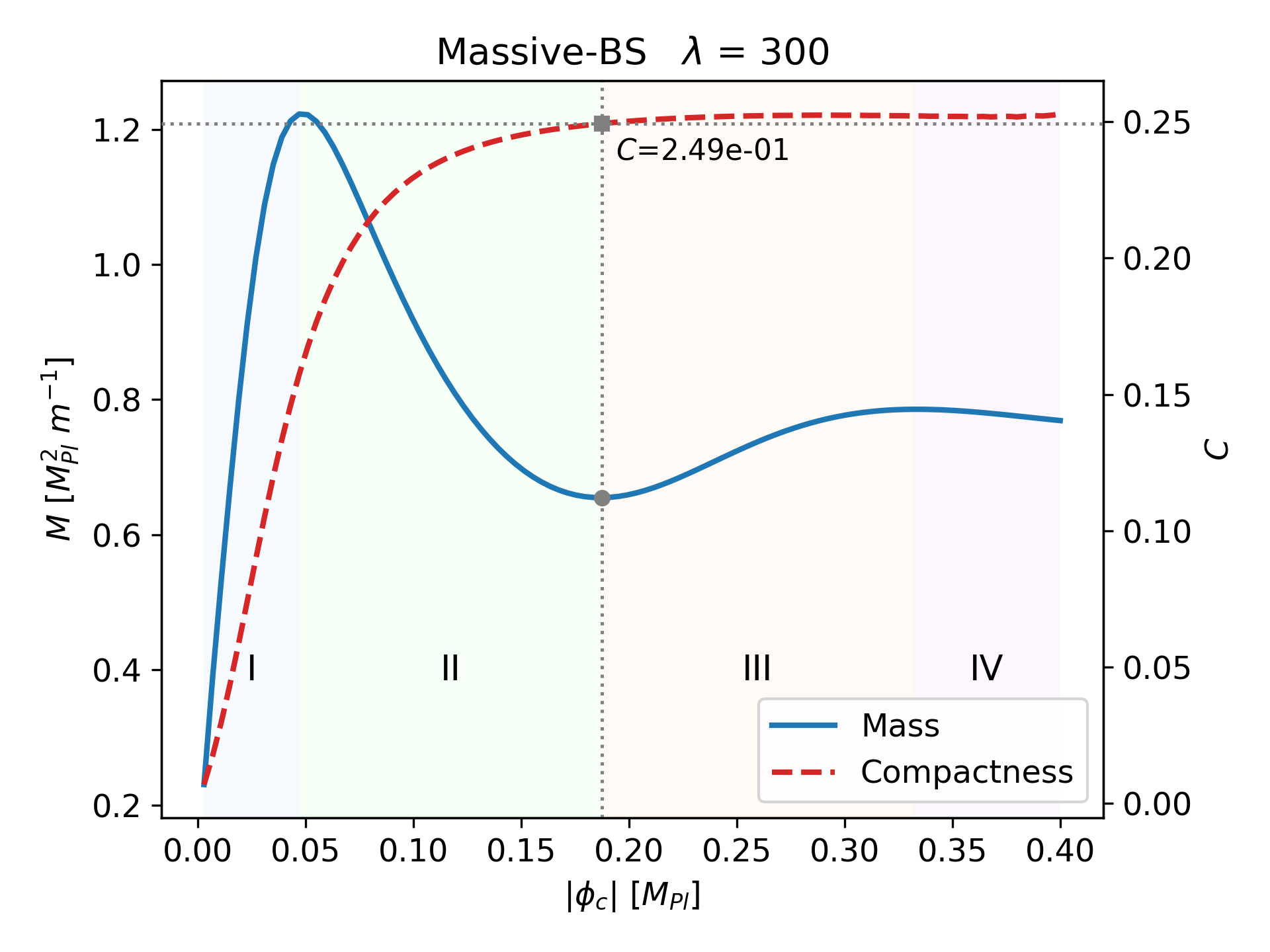}
  \end{minipage}

\caption{
Mass curves \(M(|\phi_{\rm c}|)\) and compactness curves \(\mathcal{C}(|\phi_{\rm c}|)\) for \(\lambda = 10, 50, 100,\) and \(300\). 
The extrema of the mass curves are used to divide each panel into several regions: 
region~I denotes the first segment with \(\mathrm{d}M/\mathrm{d}|\phi_{\rm c}| > 0\), 
region~II the first segment with \(\mathrm{d}M/\mathrm{d}|\phi_{\rm c}| < 0\), 
region~III the second segment with \(\mathrm{d}M/\mathrm{d}|\phi_{\rm c}| > 0\), 
and region~IV (for \(\lambda = 100, 300\)) the second segment with \(\mathrm{d}M/\mathrm{d}|\phi_{\rm c}| < 0\). 
Different regions are highlighted by different background colours. 
The numbers indicated in the panels denote the compactness \(\mathcal{C}\) at the boundary between regions~II and III.
}
\label{fig:mass_cbs}
\end{figure*}

\begin{figure*}
  \centering
  \begin{minipage}{0.49\textwidth}
    \centering
    \includegraphics[width=\linewidth]{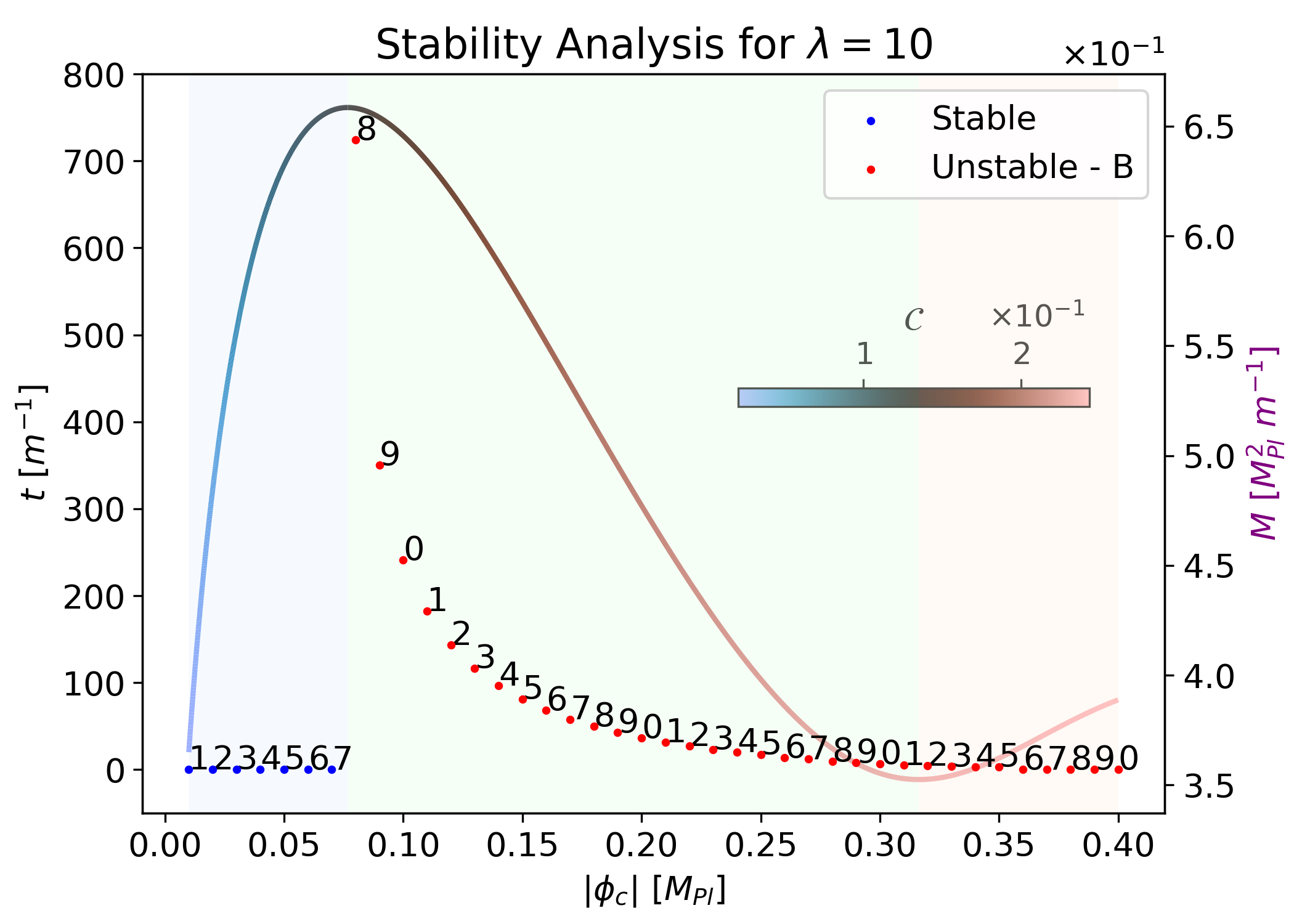}
  \end{minipage}
  \hfill
  \begin{minipage}{0.49\textwidth}
    \centering
    \includegraphics[width=\linewidth]{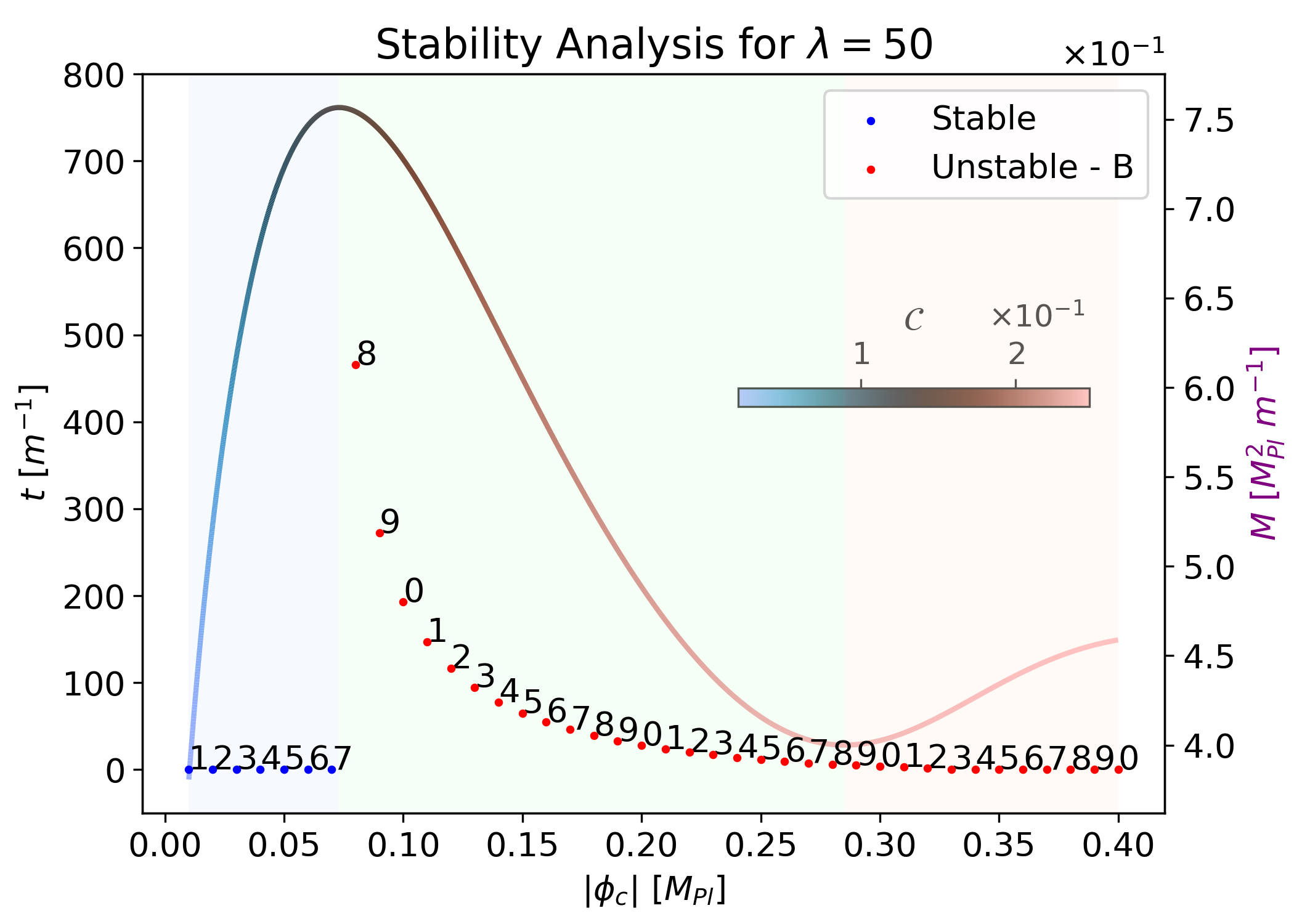}
  \end{minipage}

  \vspace{0.3cm} 

  \begin{minipage}{0.49\textwidth}
    \centering
    \includegraphics[width=\linewidth]{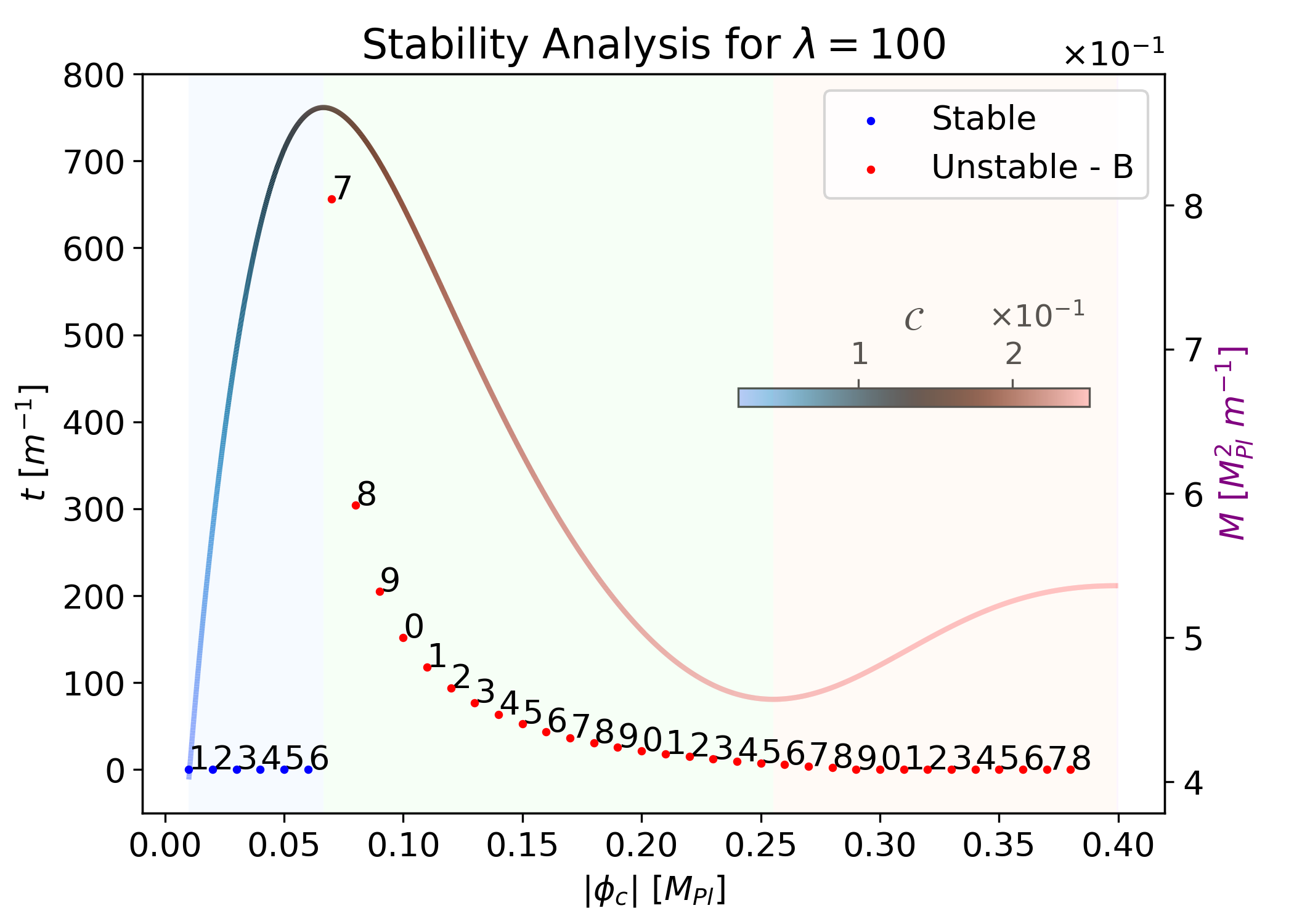}
  \end{minipage}
  \hfill
  \begin{minipage}{0.49\textwidth}
    \centering
    \includegraphics[width=\linewidth]{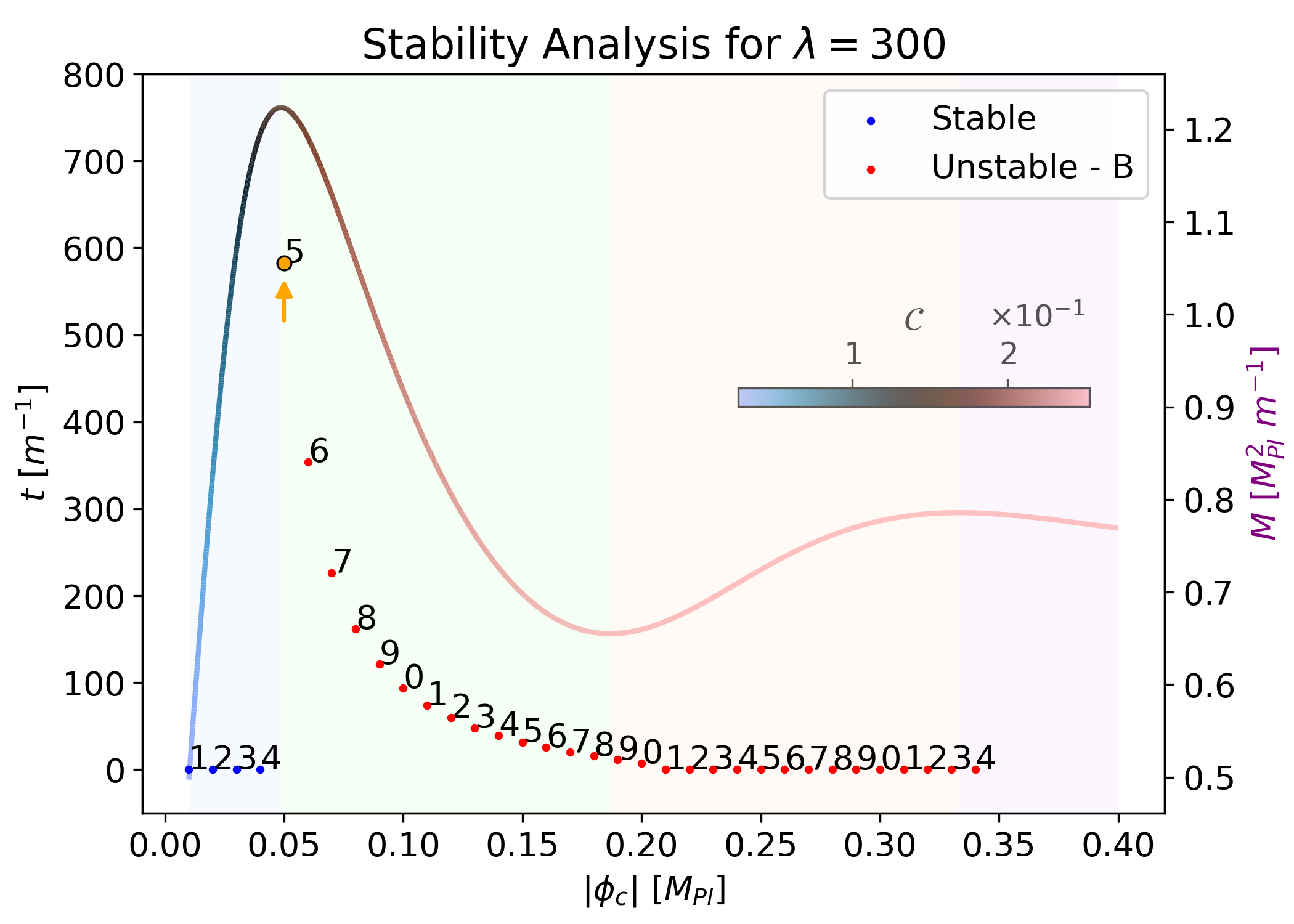}
  \end{minipage}

\caption{
The $M(|\phi_{\rm c}|)$ curves for
$\lambda = 10,~50,~100,~300$, colour--coded by the compactness $\mathcal{C}$.
The left vertical axis shows the migration/collapse time.
Coloured symbols indicate the stability class of selected configurations:
blue markers denote stable \textbf{S}-branch stars, red markers denote
\textbf{UB} models that collapse to a black hole. An orange marker indicates a special
\textbf{UB}-type boson star that collapses to a black hole after undergoing two dives.
Black-hole formation is diagnosed by monitoring the conformal factor $\chi$; for the red markers we define the collapse time as the first time at which $\min(\chi)<0.01$ and does not recover thereafter.
The vertical position of each symbol encodes the corresponding collapse or migration time; for the special orange marker the time is defined as the instant when the central amplitude $|\phi_c|$ has changed by $1\%$.
The number adjacent to each symbol indicates the last digit of the associated
$|\phi_{\rm c}|$ value.
}
\label{fig:dynamic_stability}
\end{figure*}


In the massive boson star model considered in this work, the scalar field potential is taken to be of the form \eqref{massiveV}, i.e. a polynomial potential containing only a mass term and a quartic self–interaction. For fixed \(\lambda\), the ground–state (\(n=0\)) boson stars form a one–parameter family of solutions labelled by the central amplitude \(|\phi_{\rm c}|\). The corresponding mass–central–amplitude relation \(M(|\phi_{\rm c}|)\) and compactness–central–amplitude relation \(\mathcal{C}(|\phi_{\rm c}|)\) are shown in Fig.~\ref{fig:mass_cbs}. Following the notation of Ref.~\cite{Ge:2024itl}, the extrema of \(M(|\phi_{\rm c}|)\) naturally divide the range of \(|\phi_{\rm c}|\) into several segments: regions~I and III satisfy \(\mathrm{d}M/\mathrm{d}|\phi_{\rm c}|>0\), whereas regions~II and IV satisfy \(\mathrm{d}M/\mathrm{d}|\phi_{\rm c}|<0\). In contrast to the oscillatory structure of the mass curve, within the parameter range explored in this work the compactness
\begin{equation}
\mathcal{C} \equiv \frac{M}{R} 
\end{equation}
(with $R\equiv r_{99}$ the areal radius enclosing $99\%$ of the total (ADM) mass $M$, following Ref.~\cite{Helfer:2021brt})
increases monotonically along the ground–state sequence as \(|\phi_{\rm c}|\) grows. In particular, for relatively large self–interaction strengths \(\lambda\), we find that the compactness rises very steeply across regions~I and II, reaching values as high as \(\mathcal{C}\simeq 0.24\text{--}0.25\) by the onset of region~III. 

Our dynamical simulations indicate that such highly compact configurations are extremely
sensitive to these numerically induced perturbations: even the small truncation and
constraint-violation errors present in our initial data are sufficient to trigger motion
along the unstable radial mode and lead to rapid collapse to a black hole. The stability
classification and the associated collapse or migration times of the selected
configurations are summarized in Fig.~\ref{fig:dynamic_stability}.

To highlight the differences between the massive and solitonic models in the dynamical fate of unstable branches, we adopt the same segmentation of the mass–amplitude curve as in Ref.~\cite{Ge:2024itl} and follow their branch nomenclature: stable branches are denoted by S; unstable branches whose evolution leads to black–hole formation are denoted by UB (unstable–BH); unstable branches that, after radiating away energy while approximately conserving the Noether charge \(Q\), migrate to another stable branch are denoted by UM (unstable–migrating). For suitable choices of the solitonic potential, regions~I and III contain stable configurations with moderate compactness, whereas regions~II and IV are entirely populated by unstable configurations. Time evolutions show that, for configurations in the right part of region~II, with relatively large \(|\phi_{\rm c}|\), the dominant unstable mode typically exhibits UM behaviour: during the evolution these configurations shed a small amount of mass through scalar radiation and, under the approximate conservation of the Noether charge $Q$, migrate along an almost constant--$Q$ trajectory towards larger $|\phi_c|$, eventually settling into stable configurations in region III with slightly higher compactness.

By contrast, configurations in the left part of region~II, with smaller \(|\phi_{\rm c}|\), are mostly of UB type: their mass is already ``too large'', and no stable boson star in region~III can be found that could act as a possible final state and halt the collapse. As emphasized in Ref.~\cite{Ge:2024itl}, 
``\textit{BSs starting on branch II\(_{\rm UB}\) (further left) have too large a mass to find their collapse halted by a stable BS configuration on the second bump and thus are doomed to form a BH.}''

In the massive quartic model, the mass–amplitude curve likewise exhibits multiple extrema and can be partitioned into regions~I–IV in the same formal manner. The crucial difference with respect to the solitonic case is that, within the range of \(\lambda\) considered here, the entire region~III has already entered a highly compact and dynamically unstable regime, so that no ``second stability window'' analogous to that of the solitonic potential exists. This directly alters the fate of the unstable branch in region~II: for the vast majority of massive boson stars in region~II, there are simply no accessible stable configurations at larger central amplitudes that could serve as dynamical endpoints. Under the approximate conservation of the Noether charge \(Q\), these configurations cannot migrate to a stable branch on the right as in the solitonic case; instead, the radial unstable mode drives them towards further contraction, and they promptly collapse to black holes, displaying UB–type behaviour. For \(\lambda=10,50,100\), we indeed do not observe any UM–type configurations in region~II: all unstable models ultimately collapse to black holes, in full agreement with the above picture.

\begin{figure*}
  \centering
  \begin{minipage}{0.49\textwidth}
    \centering
    \includegraphics[width=\linewidth]{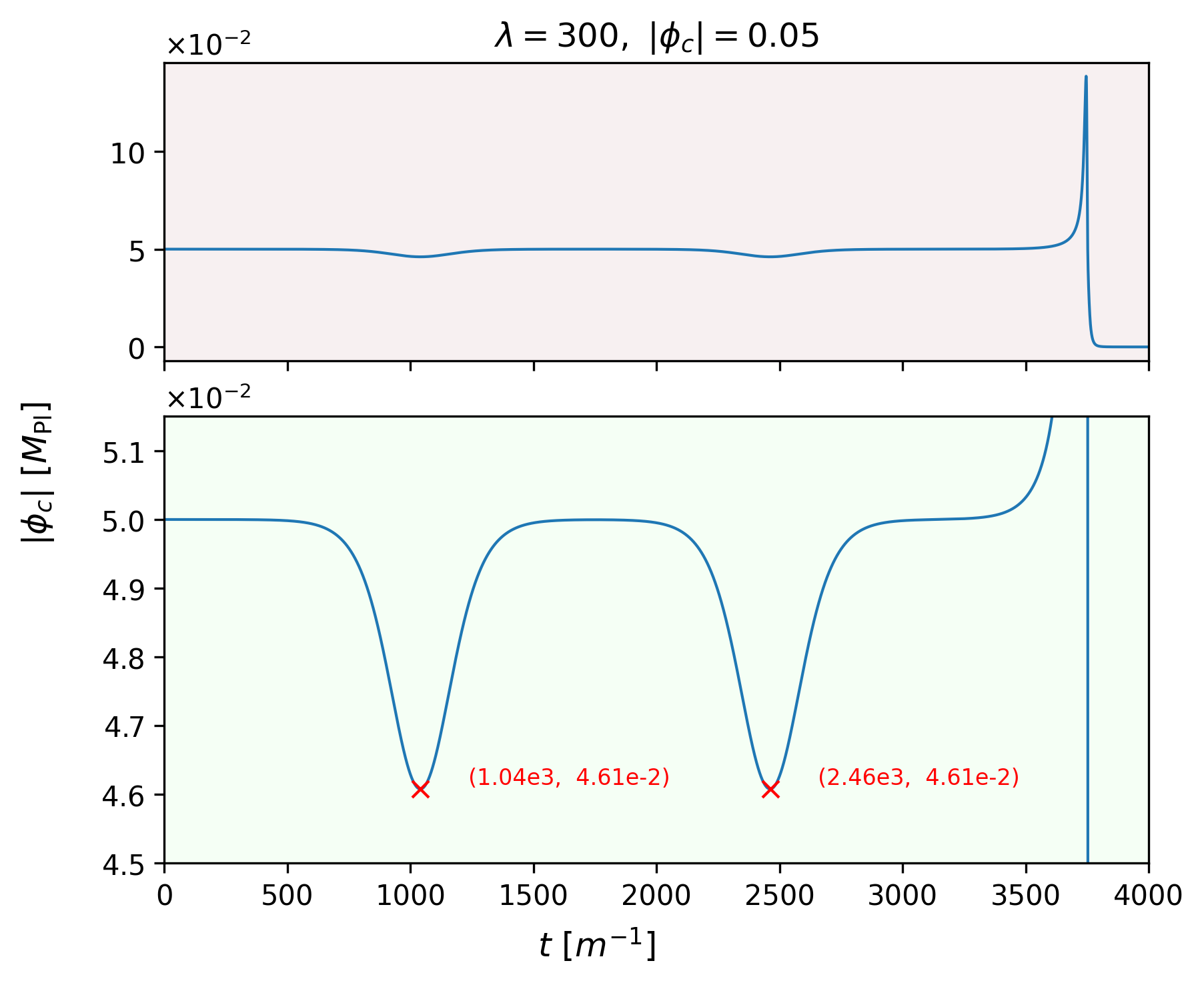}
  \end{minipage}
  \hfill
  \begin{minipage}{0.49\textwidth}
    \centering
    \includegraphics[width=\linewidth]{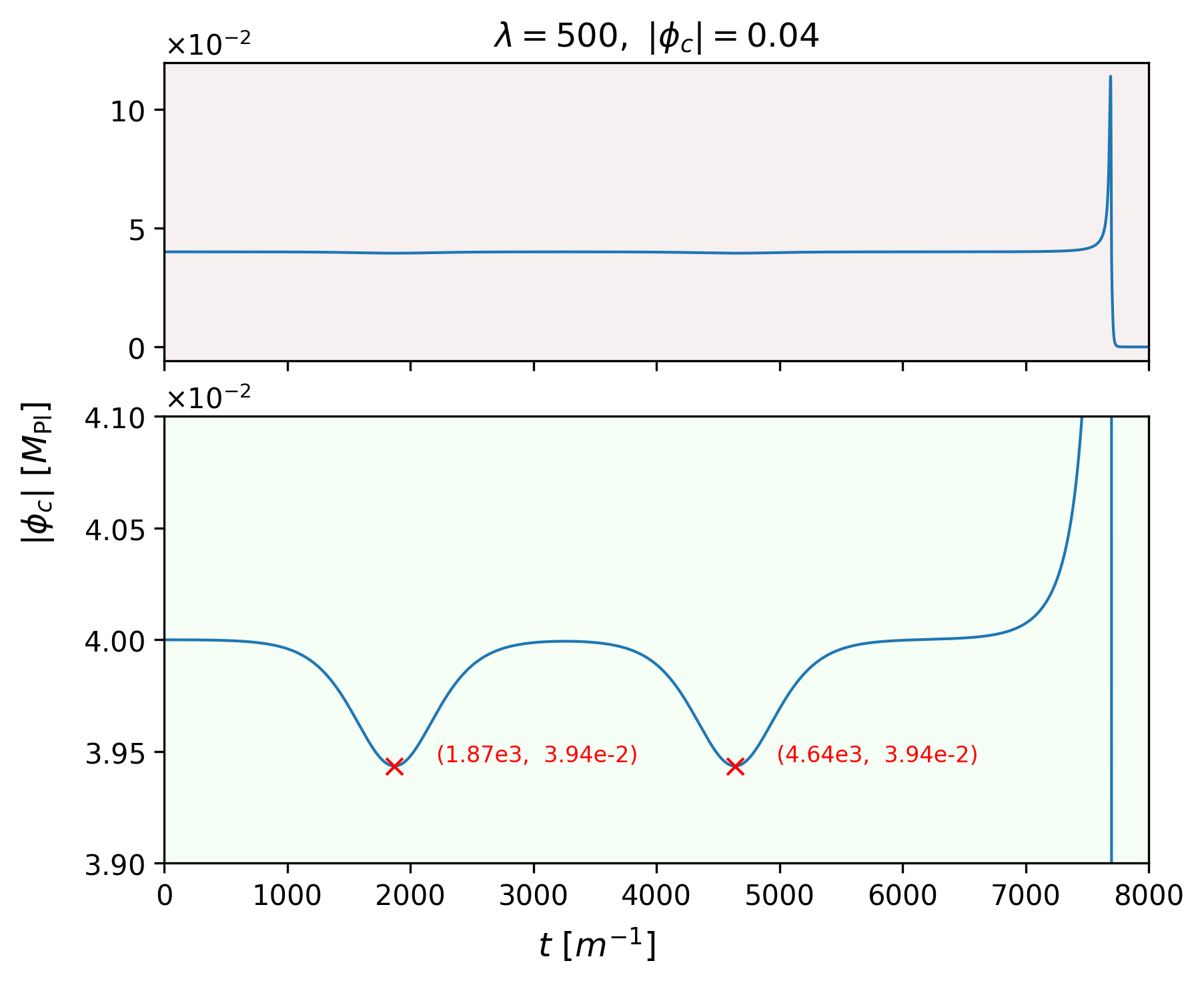}
  \end{minipage}

\caption{Evolution of the central amplitude for $\lambda = 300$, $|\phi_c| = 0.05$ (left panel), and $\lambda = 500$, $|\phi_c| = 0.04$ (right panel). For each case, the upper subplot shows the full time evolution, while the lower subplot shows the same time interval with a vertically magnified scale to highlight the ``double--dive'' behaviour. The red labels indicate the coordinates $(t,|\phi_c|)$ of the valley bottoms.}
\label{fig:dive}
\end{figure*}

An interesting case arises for $\lambda=300$. The configuration with $|\phi_{\rm c}|=0.05$, located just to the right of the first mass maximum, does not exhibit genuine UM–type migration. Instead, its time evolution shows a characteristic ``double–dive'' behaviour of the \emph{central scalar amplitude}. As displayed in the left panel of Fig.~\ref{fig:dive}, $|\phi_{\rm c} (t)|$ undergoes two downward excursions from $0.0500$ to $0.0461$, corresponding to a relative decrease of about $7.8\%$. After each dive the configuration reverses direction and returns close to its initial central amplitude rather than continuing toward smaller $|\phi_{\rm c}|$ along an approximately constant–$Q$ trajectory. The conformal factor $\chi$ remains near equilibrium during these excursions and is used only as a diagnostic to determine the onset of black–hole formation, which occurs at $t\simeq 3740$.

A similar pattern occurs for even larger self–interaction strengths. The right panel of Fig.~\ref{fig:dive} shows the example $\lambda=500$, $|\phi_{\rm c}|=0.04$. Here $|\phi_{\rm c} (t)|$ again exhibits two shallow dives, decreasing from $0.0400$ to $0.0394$, i.e. a relative drop of about $1.5\%$. These excursions are significantly smaller in amplitude, making them more difficult to resolve in practice. Black–hole formation, as diagnosed from the behaviour of $\chi$, occurs at $t\simeq 7680$. This configuration, like its $\lambda=300$ counterpart, lies immediately to the right of the first mass maximum and marks the entry into region~II.

In our parameter survey the double–dive behaviour is most clearly seen for relatively large self–interaction strengths and for models lying just to the right of the first mass maximum. We cannot, however, exclude the possibility that similar near–critical configurations exist at smaller $\lambda$ in an extremely narrow neighbourhood of the first maximum. Resolving such a region would require substantially denser sampling and higher computational cost than explored here, so we refrain from making any stronger statement about the $\lambda$–dependence of this phenomenon.

These double–dive events therefore do not represent migration toward a stable configuration. Instead, they are naturally interpreted as signatures of the unstable radial mode: small, numerically induced perturbations excite both the growing mode and oscillatory radial components, so that the central amplitude executes one or two sizeable excursions before the overall secular trend toward collapse dominates. Similar long–lived, oscillatory excursions of the central field, followed eventually by black–hole formation, have been reported in earlier studies of perturbed boson stars and black–hole threshold phenomena~\cite{Hawley:2000dt, Hawley:2000tv, Liebling:2012fv}.

\begin{figure}
\centering
\includegraphics[width=0.8\textwidth]{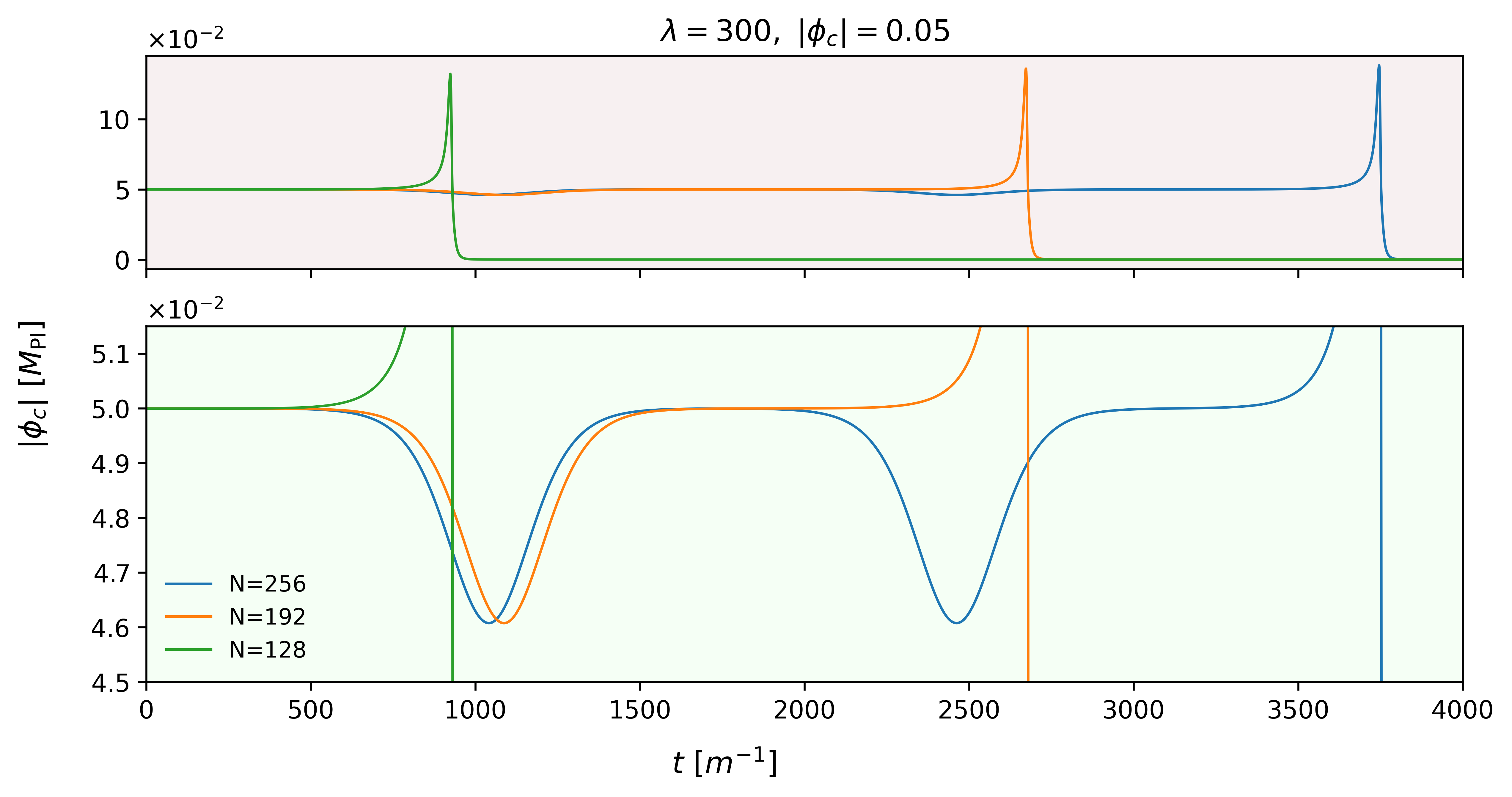}
\caption{Time evolution of the central scalar amplitude $|\phi_{\rm c}(t)|$ for the near--critical single--boson--star configuration with $\lambda=300$ and $|\phi_{\rm c}|=0.05$, shown at three resolutions ($N=128,192,256$). 
Top: full evolution; bottom: zoom--in highlighting the double--dive excursions. 
The onset of collapse occurs progressively later with increasing resolution.}\label{fig:double_dive_conv}
\end{figure}

Because these near--critical excursions are expected to be strongly sensitive to truncation error, we performed a dedicated resolution study for $\lambda=300$ and $|\phi_{\rm c}|=0.05$ (Fig.~\ref{fig:double_dive_conv}). Here and throughout, the quoted resolution $N$ refers to the level--0 (base-grid) linear resolution. All runs use adaptive mesh refinement with \texttt{max\_level}=6, i.e.\ seven refinement levels $\ell=0,\dots,6$, and a 2:1 refinement ratio between adjacent levels, $\Delta x_{\ell+1}=\Delta x_\ell/2$. The results show that the double--dive dynamics is strongly resolution dependent. At $N=128$ the evolution collapses already near the first dive. Increasing the resolution to $N=192$ allows the configuration to complete the first dive, but collapse is triggered during the second dive. At $N=256$ the evolution survives longer and the collapse point shifts again, consistent with the interpretation that insufficient resolution can induce premature collapse and alter how many dives are cleanly resolved. It is therefore plausible that the collapse locations observed at finite resolution correspond to additional ``successful dives'' that would be resolved at still higher numerical precision, before the unstable mode ultimately drives collapse.

Consequently, all region~II configurations we have evolved—including those very close to the first mass extremum—eventually collapse to black holes.

Finally, we comment on the implications of these near–critical models for the binary evolutions discussed in Sec.~\ref{sec:BSC}. For the choices of initial separation adopted in this work, the collapse and double–dive timescales of the relevant isolated configurations remain much longer than the binary contact time, so they do not qualitatively affect the head–on collisions analysed below. If one were to consider binaries with substantially larger initial separations, such that the collision timescale greatly exceeds the collapse time, the natural expectation is that each star would first collapse to a black hole and subsequently form a BH–BH binary. Conversely, if the binary timescale were tuned to be comparable to the double–dive excursions of the central amplitude, the resulting time–dependent compactness during the approach could imprint sharp, ``needle–like'' features in the emitted gravitational-wave energy, similar to those reported in Ref.~\cite{Ge:2024itl}.

\section{Boson Stars Collisions}
\label{sec:BSC}

After constructing equilibrium solutions for single boson stars, we generate
initial conditions for binary configurations by superposing two isolated
solutions with an appropriate prescription. It is important to stress that,
unlike black holes, boson stars do not possess an event horizon; consequently,
their internal dynamics can directly influence the exterior spacetime.

As shown in our earlier work~\cite{Helfer:2021brt}, a plain superposition of two
single-star configurations---although adequate for binary black holes---may
introduce non-negligible constraint violations for boson star binaries and can
lead to unphysical artefacts such as spurious or prematurely triggered
collapses. To avoid these issues, we employ the improved boson star
initial-condition construction proposed in
Refs.~\cite{Helfer:2021brt,Helfer:2018vtq}, which significantly reduces such
unphysical behaviour.

In this work we consider head-on collisions with vanishing impact parameter.
The initial separation of the stars is set to $D=80~m^{-1}$ and the initial
velocity to $v=0.1$, following the numerical setup used in
Ref.~\cite{Ge:2024itl}.

\begin{figure*}
  \centering
  \begin{minipage}{0.49\textwidth}
    \centering
    \includegraphics[width=\linewidth]{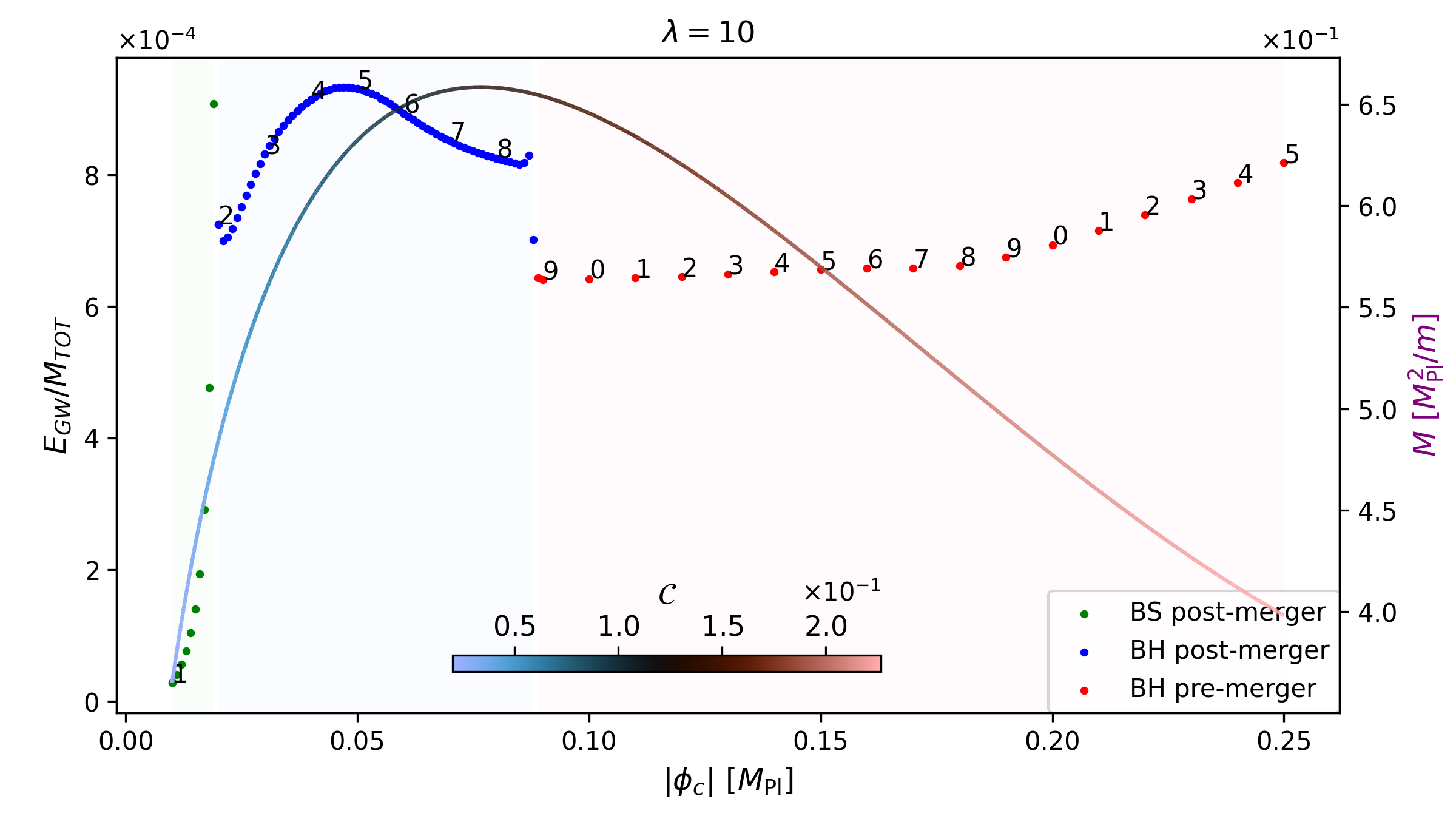}
  \end{minipage}
  \hfill
  \begin{minipage}{0.49\textwidth}
    \centering
    \includegraphics[width=\linewidth]{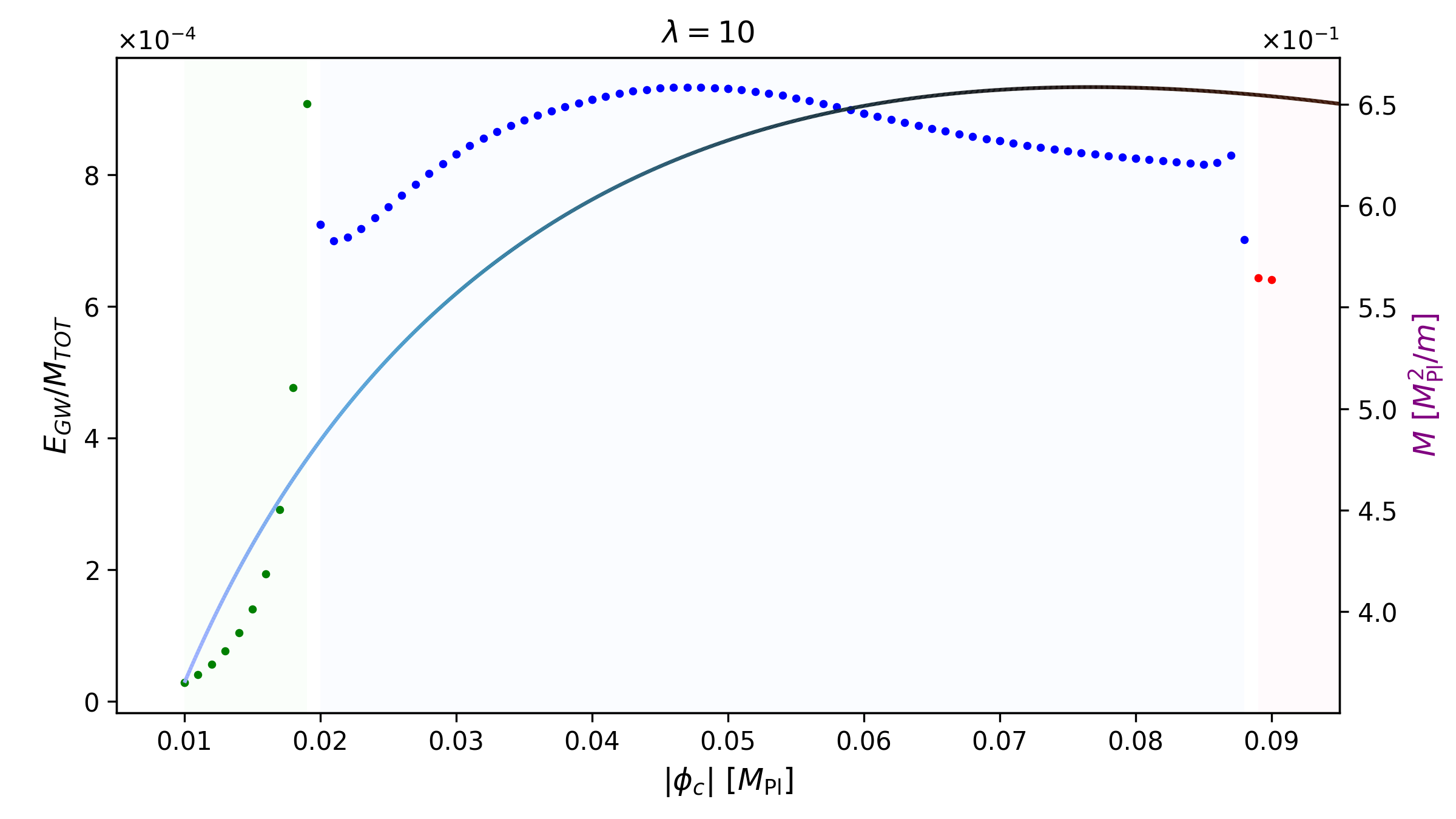}
  \end{minipage}

  \vspace{0.3cm} 

  \begin{minipage}{0.49\textwidth}
    \centering
    \includegraphics[width=\linewidth]{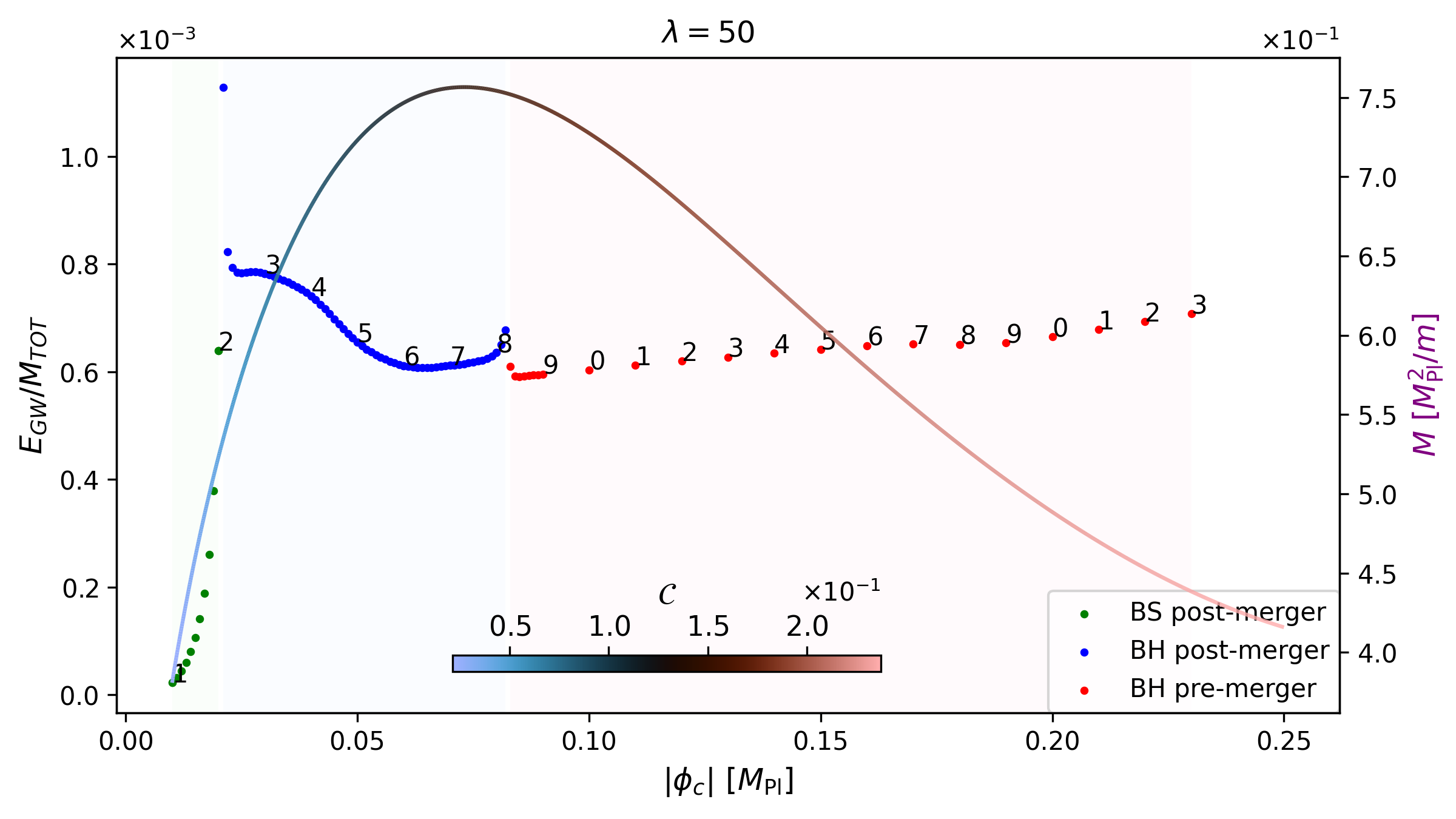}
  \end{minipage}
  \hfill
  \begin{minipage}{0.49\textwidth}
    \centering
    \includegraphics[width=\linewidth]{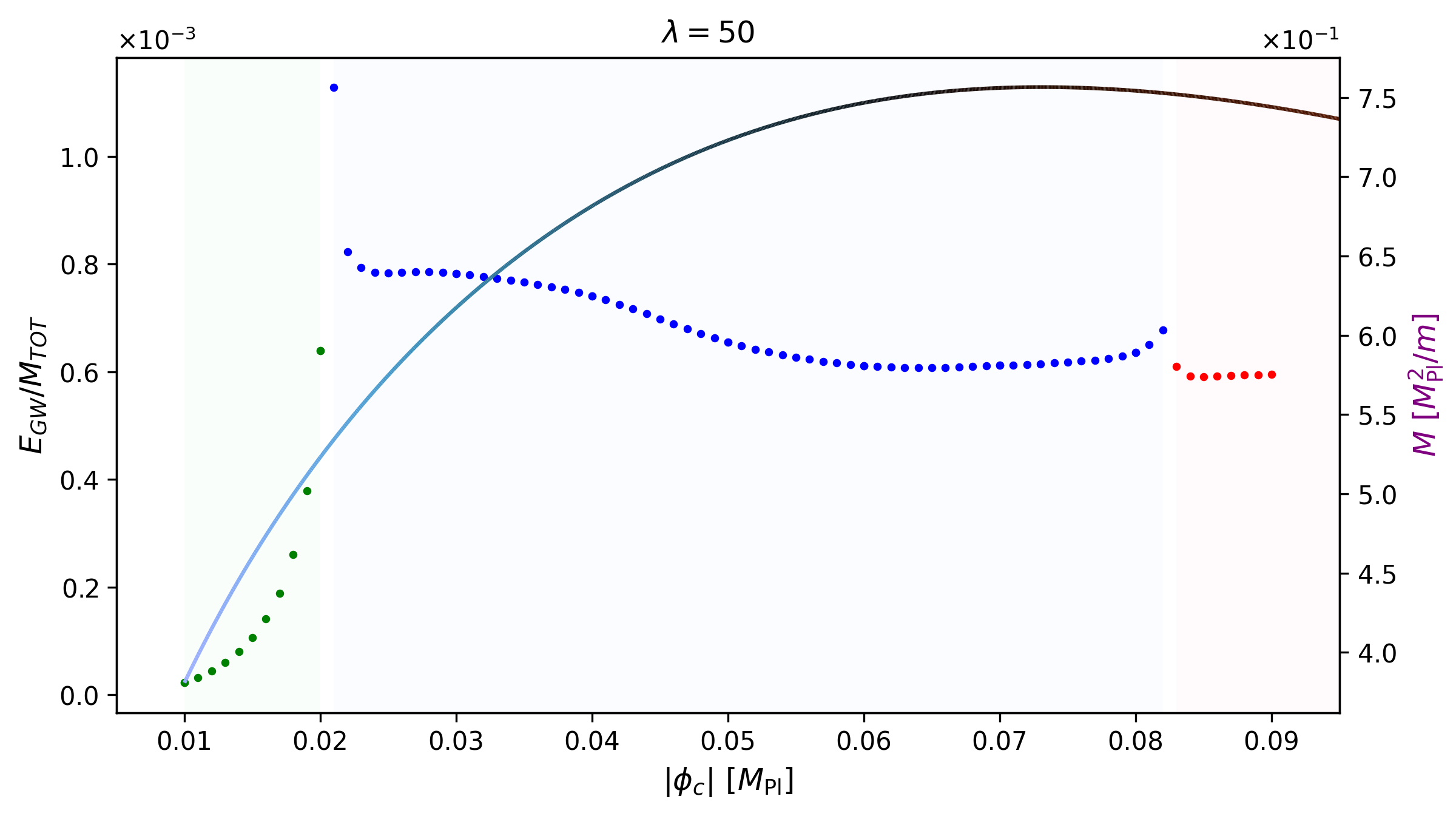}
  \end{minipage}
  
  \vspace{0.3cm} 
  
  \begin{minipage}{0.49\textwidth}
    \centering
    \includegraphics[width=\linewidth]{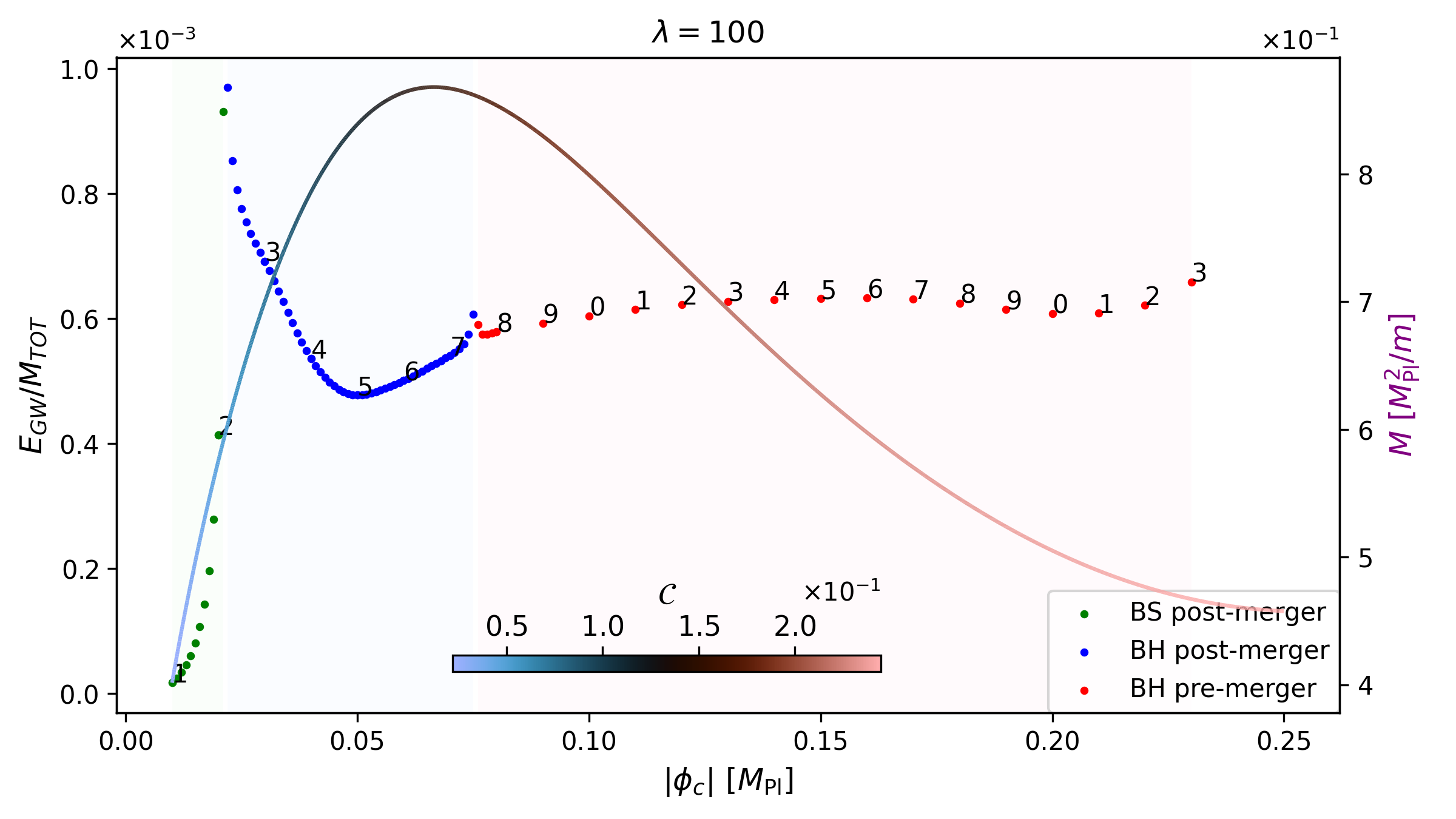}
  \end{minipage}
  \hfill
  \begin{minipage}{0.49\textwidth}
    \centering
    \includegraphics[width=\linewidth]{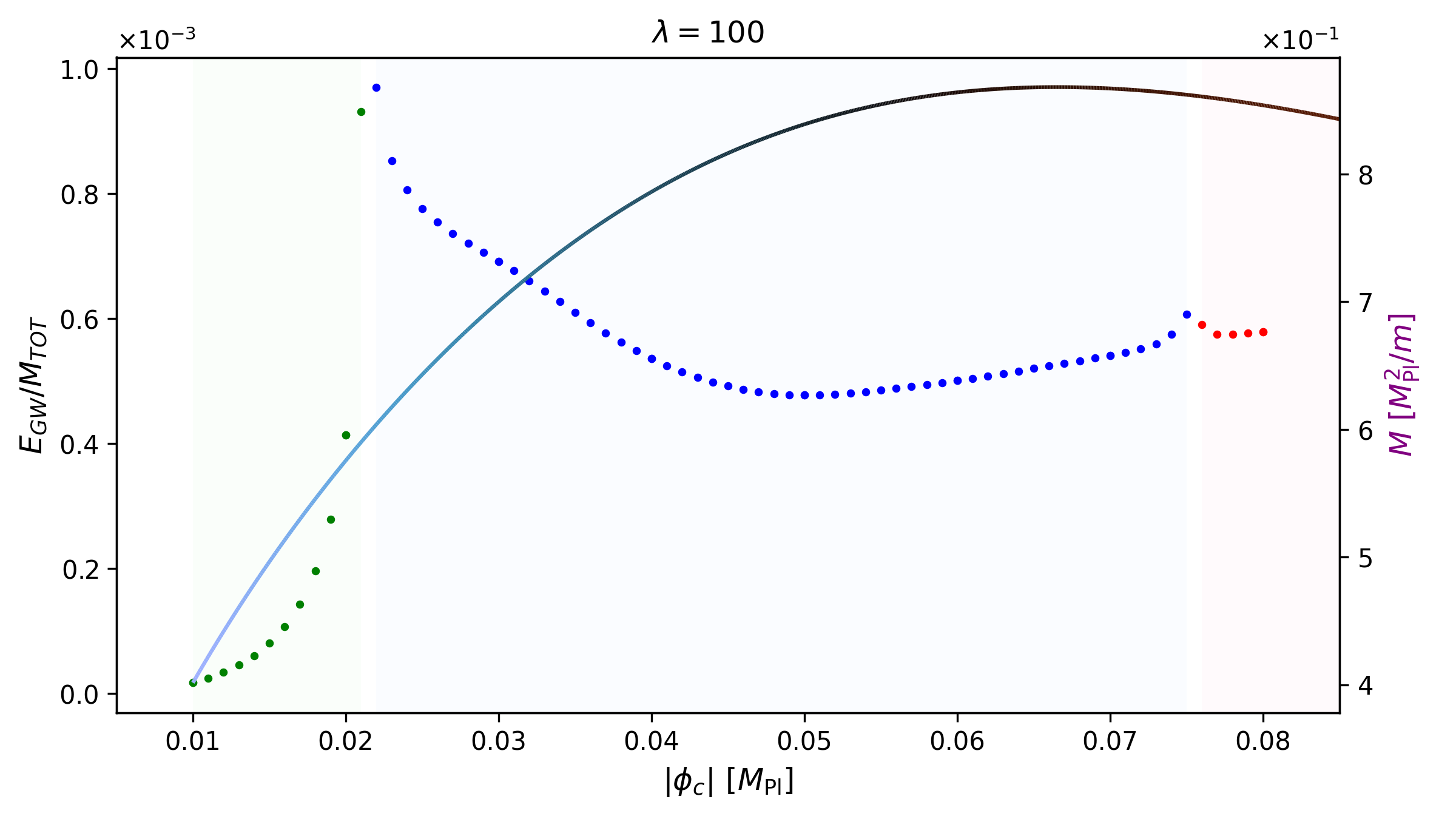}
  \end{minipage}
  
  \vspace{0.3cm} 
  
  \begin{minipage}{0.49\textwidth}
    \centering
    \includegraphics[width=\linewidth]{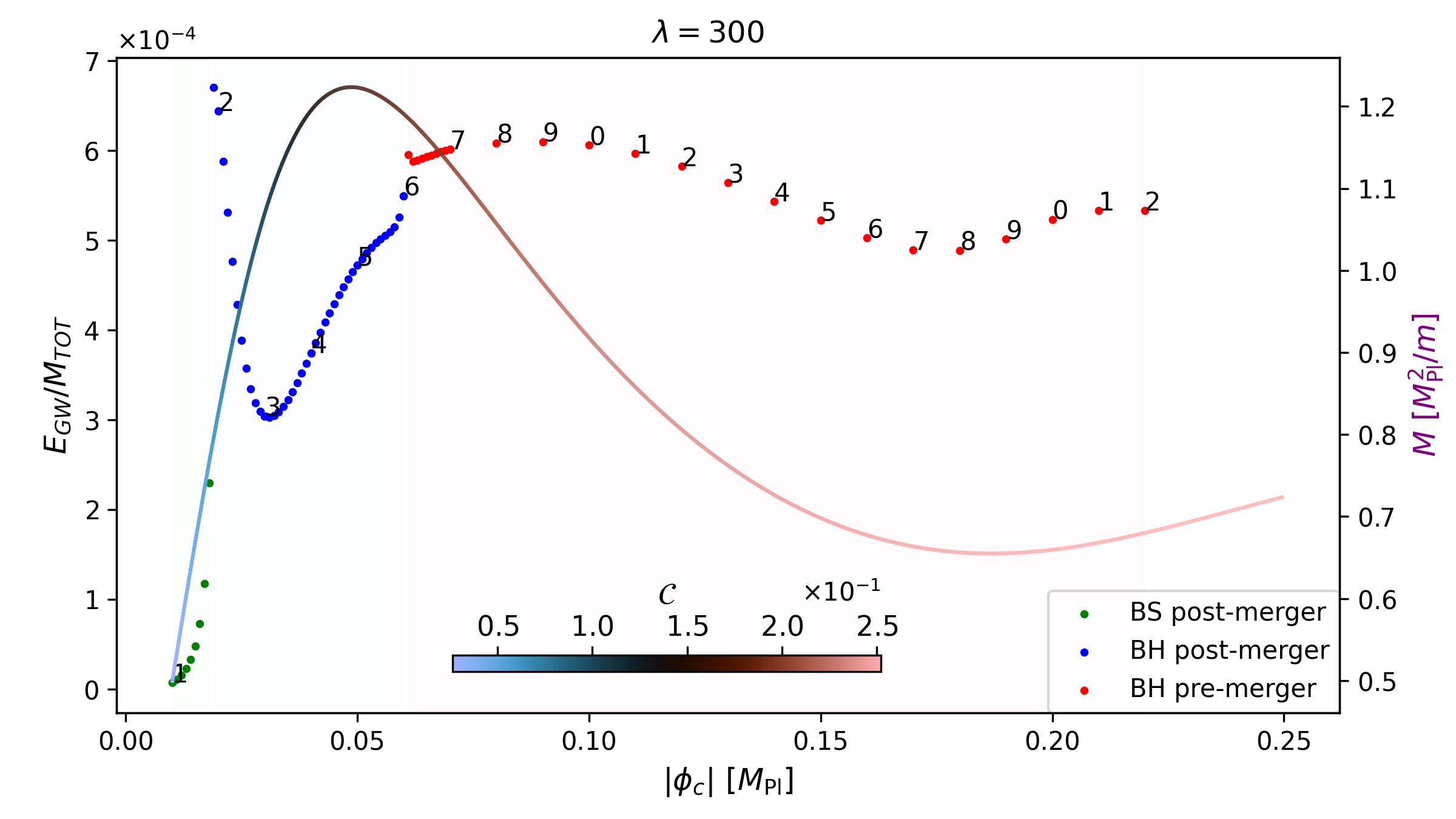}
  \end{minipage}
  \hfill
  \begin{minipage}{0.49\textwidth}
    \centering
    \includegraphics[width=\linewidth]{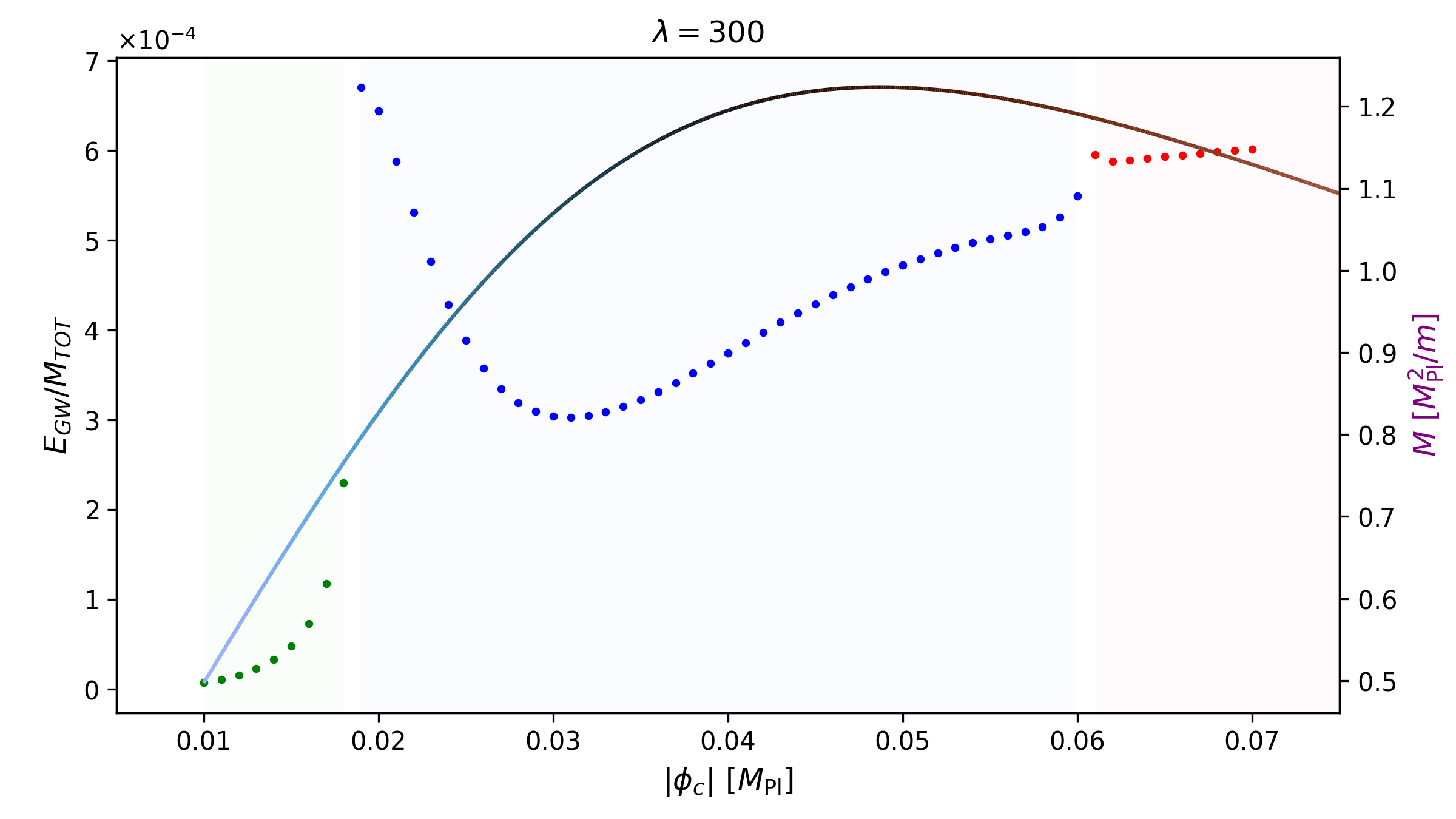}
  \end{minipage}
\caption{Head-on collisions of massive boson stars for four values of the self-coupling parameter $\lambda$. Green markers denote binaries that merge into a boson star remnant (BS post-merger), blue markers correspond to direct formation of a single black hole (BH post-merger), and red markers represent cases in which each star collapses individually prior to contact, followed by a subsequent BH--BH merger (BH pre-merger). The vertical axis shows the total emitted gravitational-wave energy, normalized by the initial total mass $M_{\rm TOT} \equiv M_1 + M_2 = 2M(|\phi_c|)$, where $M(|\phi_c|)$ is the ADM mass of a single isolated boson star with central field amplitude $|\phi_c|$, $E_{\rm GW}/M_{\rm TOT}$. The panels on the right provide zoomed-in views of the panels on the left.}
\label{fig:main_result}
\end{figure*}

Fig.~\ref{fig:main_result} summarizes our main results for the four self-interaction strengths $\lambda$ considered. Green points mark binaries that
merge into a boson star remnant (BS post-merger), blue points represent cases in which the merger directly forms a black hole (BH post-merger), and red points correspond to evolutions in which the two boson stars individually collapse into black holes prior to contact, followed by a BH--BH coalescence (BH pre-merger).

In our previous work~\cite{Ge:2024itl}, we pointed out that the magnitude of the gravitational-wave energy is governed by a competition between two effects: the increase in compactness, which tends to enhance the efficiency of gravitational-wave emission, and the reduction of tidal deformability, which suppresses merger asymmetries and thus tends to reduce the emitted radiation. Consequently, even less compact boson stars may in some cases radiate more strongly if they remain sufficiently deformable to develop larger asymmetries during merger. This qualitative picture is also confirmed by the results presented here for massive boson stars.

\subsection*{BS/BH post-merger Region~I }

This region contains two types of merger outcome for boson star binaries: BS
post-merger and BH post-merger. In both cases, the qualitative behaviour is
similar to that of solitonic boson stars: as $|\phi_c|$ increases, the
normalized gravitational-wave energy first rises and then decreases. The
difference is that, for massive boson stars, the turning point associated with
the ``optimal compactness'' occurs at smaller $|\phi_c|$. As shown in
Fig.~\ref{fig:mass_cbs}, the compactness of massive boson stars increases more
steeply with $|\phi_c|$ than in the solitonic case, so that the turning point is
reached earlier along the sequence.

\subsection*{BH post-merger Region~II }

Upon entering Region~II, the massive boson stars lie on the unstable branch, characterized by $dM/d|\phi_c|<0$. In this regime, the fate of the binary is controlled by the competition between the collapse timescale of the individual stars and the binary contact time. In this subsection we focus on those models for which the collapse time is \emph{longer} than the contact time (typically $t \gtrsim 300$ in our simulations). In these cases the stars still undergo a head-on collision before collapse, and the outcome is a BH post-merger configuration; these systems correspond to the blue markers in Fig.~\ref{fig:main_result}. The corresponding waveforms and radiated energies retain many of the qualitative features already seen in solitonic boson star mergers.

It is worth noting that, as $\lambda$ is increased, the blue points in
Region~II develop a second upward trend in $E_{\rm GW}/M_{\rm TOT}$, a feature that is particularly pronounced for $|\phi_c|\simeq 0.03$–$0.07$ in the $\lambda=300$ sequence. We interpret this behaviour as a manifestation of the rapidly increasing
compactness in this interval: the rise in compactness becomes strong enough to dominate over the loss of asymmetry, thereby enhancing the efficiency of gravitational-wave emission.

\subsection*{BH pre-merger}

One of the most striking differences with Ref.~\cite{Ge:2024itl} occurs in the regime where the individual boson stars collapse \emph{before} coming into contact. Whenever the collapse time becomes shorter than the binary contact time (approximately $t\simeq 300$ in our setup), each star collapses to a black hole prior to merger, and the late-time evolution reduces to a BH--BH head-on collision. These configurations are represented by the red markers in Fig.~\ref{fig:main_result}.

In our previous study~\cite{Ge:2024itl}, the BH post-merger channel was found to be systematically louder than the BH pre-merger channel. In the quartic massive boson star model, this hierarchy persists only for small values of $\lambda$: as $\lambda$ increases,
the gravitational-wave energies of the BH post-merger and BH pre-merger channels become comparable, 
and for $\lambda=300$ the average level of the BH pre-merger branch even exceeds that of many BH post-merger cases.

For smaller self-couplings ($\lambda = 10$ and $50$), the gravitational-wave energy along the red branch in Region~II increases monotonically with $|\phi_c|$. This behaviour is consistent with the comparatively mild growth of the compactness in this part of the sequence. In this regime, higher compactness leads to a more prompt and cleaner collapse of each star, so that a larger fraction of the initial mass ends up in the two pre-merger black holes. The subsequent BH--BH merger therefore more closely resembles a vacuum binary black hole coalescence and radiates gravitational waves more efficiently.


For larger self-couplings ($\lambda = 100$ and $300$), once $|\phi_c|$ enters a
high-compactness regime in which the compactness of the equilibrium boson stars
has essentially saturated, the BH pre-merger branch develops a clear
non-monotonic structure in the normalized gravitational-wave energy. This
behaviour is visible in the red markers of the $\lambda = 100$ and
$\lambda = 300$ panels in Fig.~\ref{fig:main_result}, and it cannot be
explained solely by the mild residual variation in compactness over this range.

\begin{figure*}
  \centering
  \begin{minipage}{0.45\textwidth}
    \centering
    \includegraphics[width=\linewidth]{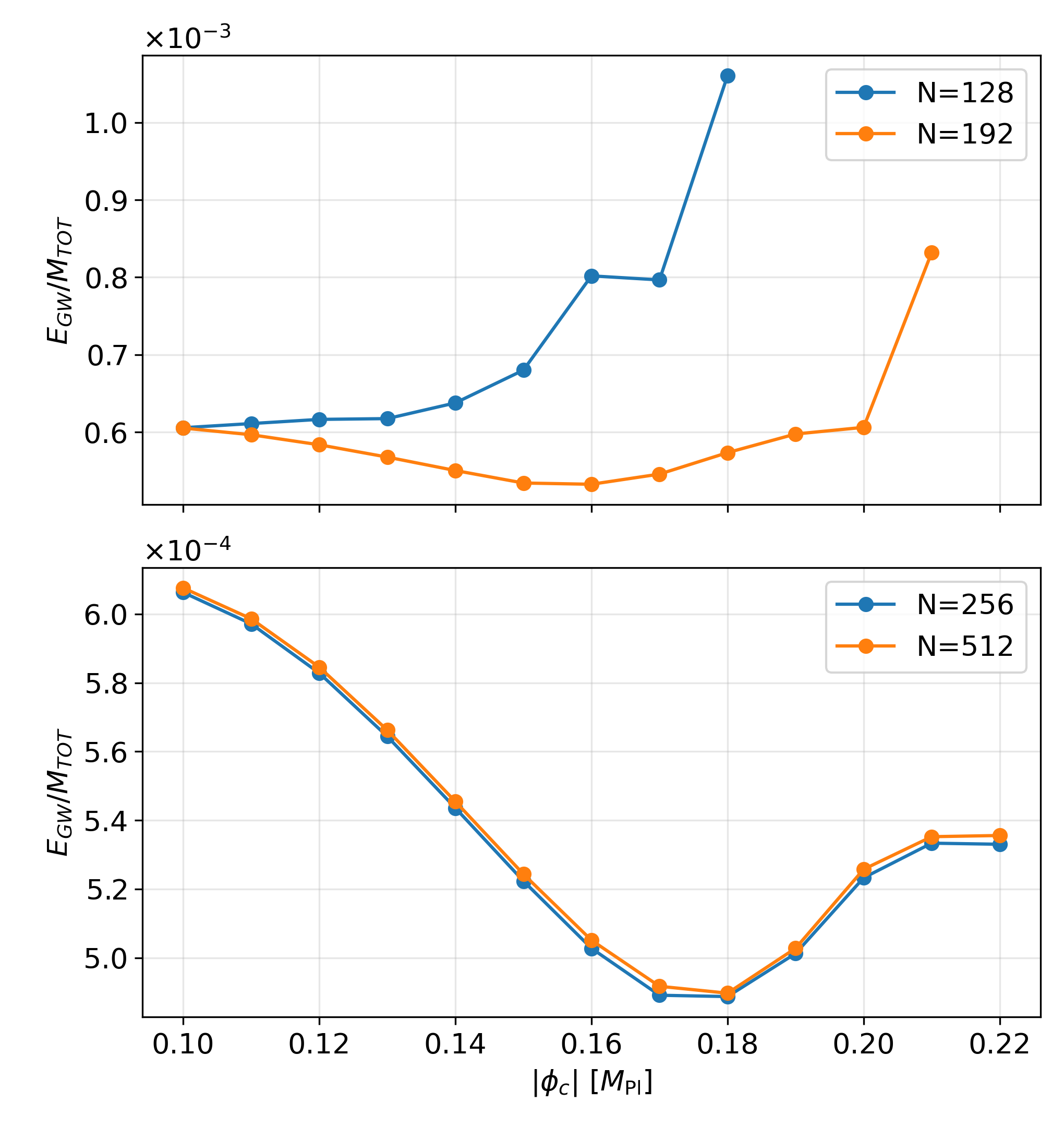}
  \end{minipage}
  \hfill
  \begin{minipage}{0.49\textwidth}
    \centering
    \includegraphics[width=\linewidth]{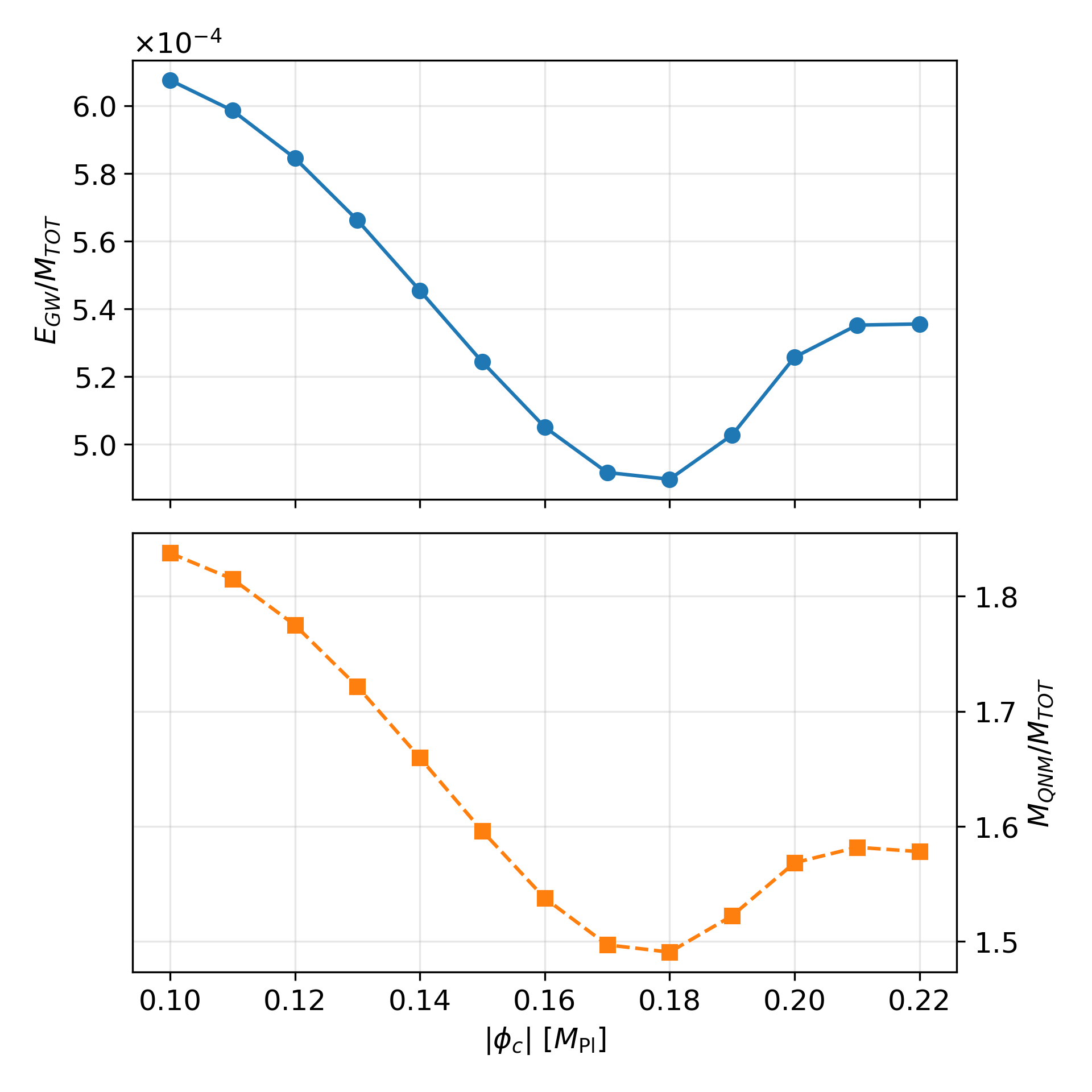}
  \end{minipage}

\caption{High--compactness regime at $\lambda=300$. \emph{Left:} $E_{\rm GW}/M_{\rm TOT}$ versus $|\phi_{\rm c}|$ at different resolutions; the non--monotonic trend is resolved for $N\ge256$ and remains stable up to $N=512$. \emph{Right:} Ringdown (QNM) mass $M_{\rm QNM}/M_{\rm TOT}$ inferred from the post--merger waveforms,
estimated from the $(\ell,m)=(2,0)$ mode of the extracted signal following the procedure of Ref.~\cite{Stein:2019mop},
showing a positive correlation with $E_{\rm GW}/M_{\rm TOT}$.}
\label{fig:HiCBS}
\end{figure*}

To verify that the high-$\lambda$ trend is numerically robust, we performed a
dedicated resolution study in the region $\lambda=300$ and $|\phi_{\rm c}|>0.1$.
As shown in Fig.~\ref{fig:HiCBS} (left), at low resolution the extracted
$E_{\rm GW}/M_{\rm TOT}$ becomes strongly resolution dependent at large
$|\phi_{\rm c}|$ and shows no sign of convergence, so the apparent trend in this
region is not numerically robust.
By contrast, the non-monotonic trend discussed above becomes clearly visible
once the resolution reaches $N=256$, and it remains stable when further
increasing the resolution to $N=512$.
To be conservative, we do not attempt to attribute the observed non-monotonicity
in $E_{\rm GW}/M_{\rm TOT}$ to a single physical mechanism. Several effects may,
in principle, contribute to the non-monotonicity: (i) even after the compactness
has saturated, small differences in the outer scalar profile of the equilibrium
solutions may influence how cleanly the individual collapses proceed; (ii)
near-critical boson star configurations are known to be highly sensitive to small
perturbations introduced by the binary superposition; and (iii) nonlinear
couplings between the early collapse dynamics and the subsequent BH--BH merger
may also play a role. 

As an additional diagnostic, Fig.~\ref{fig:HiCBS} (right) shows that
$M_{\rm QNM}/M_{\rm TOT}$ exhibits a positive trend with $E_{\rm GW}/M_{\rm TOT}$ in this regime. Here $M_{\rm QNM}$ is inferred from the ringdown of the $(\ell,m)=(2,0)$ mode of the extracted signal following the procedure of
Ref.~\cite{Stein:2019mop}; we simply report this correlation and do not
interpret it further here. With the diagnostics available in this work, we
cannot disentangle the relative importance of these possible contributions.


For the purposes of this work, the key result is that, in the
high-compactness regime, the BH pre-merger branch displays a consistent and
clearly identifiable non-monotonic behaviour at sufficiently large $\lambda$
(in our dataset, $\lambda=100$ and $\lambda=300$), and that this trend cannot be
explained solely by the nearly flat compactness profile in this region.
A more detailed decomposition of the
underlying effects---for example, by varying the initial separation, employing constraint-satisfying binary initial data, or explicitly tracking scalar-field energy fluxes---is left to future work.

\clearpage
\section{Summary}
\label{sec:summary}

In this work we extend our previous study of mini and solitonic boson
stars~\cite{Ge:2024itl} to quartically self–interacting massive boson stars,
combining equilibrium calculations with numerical relativity evolutions in the
Cartoon framework implemented in {\sc GRChombo}. Within the model defined by the
potential in Eq.~\eqref{massiveV}, we construct ground–state sequences
parametrized by the central scalar amplitude $|\phi_{\rm c}|$ for several
values of the self–interaction strength $\lambda$, and analyse both their
global properties and dynamical stability.

For isolated stars, our results indicate that quartically self–interacting
massive boson stars follow the same qualitative stability pattern recently
identified in Ref.~\cite{Santos:2024vdm} for other bosonic models. The boson
star mass $M(|\phi_{\rm c}|)$ develops multiple extrema along the sequence, but
dynamical stability changes only at the \emph{first} maximum; subsequent extrema
lie entirely within a dynamically unstable portion of the sequence. In the
parameter range explored here, the compactness $\mathcal{C}$ increases
monotonically with $|\phi_{\rm c}|$ and already reaches relatively large values,
$\mathcal{C}\simeq 0.24$–$0.25$, near the onset of region~III. Our numerical
evolutions show that these highly compact configurations are extremely sensitive
to the small, unavoidable perturbations associated with truncation errors and
residual constraint violations: even in the absence of any explicit
perturbation, these numerical effects are sufficient to excite the unstable
radial mode and trigger rapid collapse to a black hole. Within the parameter
space explored here, we find no evidence for a second stability window analogous
to the solitonic case.

In addition, for relatively large self–interaction strengths we identify
near–critical configurations just to the right of the first mass maximum, in
which the central scalar amplitude exhibits a characteristic ``double–dive''
pattern before collapse. In these models the central field undergoes one or two
sizeable excursions away from its initial value and then returns close to it,
before the secular trend toward collapse eventually dominates. This behaviour is
naturally interpreted as the result of small, generic perturbations exciting
both the unstable radial mode and oscillatory radial components. Similar
long–lived, oscillatory excursions of the central field, followed eventually by
black hole formation, have been reported in earlier studies of perturbed boson
stars and black hole threshold
phenomena~\cite{Hawley:2000dt,Hawley:2000tv,Liebling:2012fv}. For the parameter
choices explored here, however, all region~II configurations we evolve – including
those very close to the first mass maximum – ultimately collapse to black holes,
and the double–dive behaviour is clearly resolved only along the high–$\lambda$
sequences; at smaller $\lambda$ we cannot rule out similar near–critical
solutions confined to an extremely narrow neighbourhood of the first mass maximum,
but resolving such a region would require much denser sampling than considered in
this work.

We have also investigated head–on collisions of equal–mass massive boson stars
for several values of $\lambda$, using binary initial conditions constructed by
superposing single–star equilibrium solutions with the improved prescription of
Ref.~\cite{Helfer:2021brt,Helfer:2018vtq}. Depending on the location of the individual models
along the sequence, the mergers may lead to a boson–star remnant (BS post–merger),
direct black hole formation at contact (BH post–merger), or a regime in which
both stars collapse individually prior to contact and the late–time dynamics
reduce to a head–on black–hole binary (BH pre–merger). The overall trends in the
normalized gravitational–wave energy $E_{\rm GW}/M_{\rm TOT}$ can be
qualitatively understood as the outcome of a competition between two effects
already highlighted in Ref.~\cite{Ge:2024itl}: the increase in compactness,
which tends to enhance the efficiency of gravitational–wave emission, and the
reduction of tidal deformability, which suppresses merger asymmetries and thus
reduces radiation.

For small and moderate self–interaction strengths the BH post–merger channel
remains the loudest source of gravitational waves, closely mirroring the
behaviour found in the solitonic case. At larger $\lambda$, however, the
picture becomes richer. In the high–compactness regime the BH pre–merger branch
develops a clear non–monotonic structure in $E_{\rm GW}/M_{\rm TOT}$, and for
$\lambda=300$ the average radiated energy along this branch even exceeds that of
many BH post–merger configurations. This behaviour cannot be explained by the
nearly flat compactness profile alone and signals the presence of additional
dynamical effects associated with the interplay between collapse timescales,
near–critical behaviour and binary interaction dynamics.

\section{Outlook}
\label{sec:outlook}

The results presented here raise several questions that we leave for future work. 

Although the Cartoon method substantially reduces the computational cost of axisymmetric simulations and we have pushed the grid resolution as far as practical, our study remains subject to two intrinsic limitations. On the one hand, despite continued advances in computational hardware (often summarized under “Moore’s law”), practical limits on resolution and parameter-space coverage remain unavoidable, so both the achievable resolution and the extent of the parameter space that can be scanned are inherently restricted. On the other hand, our binary initial data, even when constructed using the improved superposition method of Ref.~\cite{Helfer:2021brt}, inevitably contain a certain level of constraint violation. This is a structural limitation of any superposition-based construction: the constraints are not exactly satisfied by design, so violations can be reduced but not completely removed. In principle, one could instead construct fully constraint-satisfying binary data by directly solving the constraint equations. In practice, however, this comes at a substantial computational cost and introduces additional numerical
uncertainties, so it does not automatically guarantee a clear advantage over the improved superposition method used here.

These considerations motivate the search for an approach that can, at least partially, sidestep both hardware constraints and the mathematical limitations of current initial-data constructions. Recent work by Helfer \textit{et al.}~\cite{Helfer:2024lrg} shows that neural-network techniques can be integrated into boson star modelling in a controlled way. In a similar spirit, we plan to explore AI-based strategies to refine binary initial data and to enhance the effective accuracy of numerical-relativity simulations even at comparatively modest grid resolutions. 

Meanwhile, the simulations reported here, together with our earlier study of mini and solitonic boson stars~\cite{Ge:2024itl}, already constitute a sizeable catalogue of gravitational-wave signals from head-on boson star collisions, combined with a large set of corresponding initial configurations. It is therefore natural to regard these data as training material for neural networks aimed at learning the mapping from physical parameters (such as the self-interaction strength, central amplitude, and merger channel) to waveform features and summary quantities. Developing AI-based surrogate models on top of these numerical data would enable rapid generation of approximate waveforms across a large parameter space and provide a complementary route to building template banks for future searches targeting boson star mergers. 

In this sense, the present work can be viewed not only as a step toward understanding the dynamics of massive boson stars, but also as a starting point for our future programme of AI-assisted gravitational-wave modelling of compact objects and AI-guided improvements to binary boson star simulations.

\begin{acknowledgments}
The author gratefully acknowledges HIAS for providing access to the ``Quantum Universe Physical Simulation Platform'', a supercomputing resource that was crucial for carrying out the simulations presented in this work.
The author also thanks Eugene Lim, Ulrich Sperhake, Tamara Evstafyeva, Daniela Cors, and Gareth Marks for helpful discussions.
This work was supported by the National Natural Science Foundation of China under Grant No.~12505066.
\end{acknowledgments}


\clearpage
\appendix
\section{2D Boson Star Matter Field and Evolution Equations}
\label{app:2dbsequ}


In this appendix, we present the explicit $2$D expressions for the boson star matter sector and the corresponding derivation of its evolution equations.

For completeness, we briefly summarize the conventions used in the $2$D (Cartoon) reduction. We evolve the system on the $z=0$ hypersurface in Cartesian coordinates $(x,y,z)$ while imposing axisymmetry about the $x$-axis. Under this symmetry, all tensor components are represented on the evolution plane and the dependence on the suppressed azimuthal direction around the $x$-axis is reconstructed by rotations in the $(y,z)$ plane. As a result, first derivatives with respect to $z$ vanish on $z=0$ for scalar quantities, whereas certain second-derivative combinations remain finite and are replaced by standard Cartoon identities (e.g., terms of the form $(\partial_y \cdot)/y$). This is the origin of the characteristic ``$1/y$'' contributions that appear in the reduced evolution equations.

With these conventions, the matter sector enters the spacetime evolution only through the usual $3+1$ projections of the stress--energy tensor. We therefore start by defining $\rho$, $S_\alpha$, and $S_{\alpha\beta}$ in terms of $T_{\mu\nu}$, and then provide their explicit expressions for the complex scalar field used in this work, written in a form directly implementable in the $2$D BSSN/CCZ4 code.

The matter terms in Eqs.~\eqref{eq:2dbsequ} are defined by

\begin{equation}
  \rho = T_{\mu\nu}n^{\mu}n^{\nu}\,,~~~
  S_{\alpha} = -\gamma^{\nu}{}_{\alpha} T_{\mu\nu} n^{\mu}\,,~~~
  S_{\alpha\beta} = \gamma^{\mu}{}_{\alpha} \gamma^{\nu}{}_{\beta}
        T_{\mu\nu}\,,~~~
  \gamma^{\mu}{}_{\alpha}=\delta^{\mu}{}_{\alpha}+n^{\mu}n_{\alpha}\,.
\end{equation}

Here $\rho$ is the energy density measured by an observer normal to the spatial hypersurface, $S_\alpha$ is the corresponding momentum density (energy flux) projected onto the hypersurface, and $S_{\alpha\beta}$ is the spatial stress tensor. Once $T_{\mu\nu}$ is specified for the matter model, these projections follow algebraically from the above definitions.

In our implementation, the complex scalar field is evolved by decomposing it into its real and imaginary parts,
\begin{equation}
\phi = \phi_{\mathrm{Re}} + i \phi_{\mathrm{Im}},
\end{equation}
and by introducing the standard first-order-in-time variable
\begin{equation}
\Pi = \Pi_{\mathrm{Re}} + i \Pi_{\mathrm{Im}},
\end{equation}
used in the main evolution system. Substituting the corresponding stress--energy tensor $T_{\mu\nu}$ into the projections above yields explicit matter source terms in a form directly compatible with the 2D Cartoon reduction, including the geometric contributions that arise from enforcing axisymmetry (such as the characteristic $1/y$ terms).

Below, we present the expressions for these matter terms within the BSSN/CCZ4 framework of the 2D code.

\subsection*{Evolution of $\phi$}

We begin with the standard $3+1$ evolution equation for the scalar field,
$\partial_t \phi = \beta^I \partial_I \phi - \alpha\,\Pi$, and then apply the
$2$D Cartoon reduction on the $z=0$ plane. With the index split $I=(i,a)$,
where $i\in\{x,y\}$ lies in the evolution plane and $a=z$ denotes the suppressed
direction, axisymmetry implies $\partial_a \phi = 0$ for scalar quantities on
$z=0$, so that the $a$-contribution drops out.

\begin{equation}
\begin{aligned}
\partial_t \phi_{\mathrm{Re}}
= \beta^I \partial_I \phi_{\mathrm{Re}} - \alpha\, \Pi_{\mathrm{Re}} 
= \beta^i \partial_i \phi_{\mathrm{Re}} + \beta^a \partial_a \phi_{\mathrm{Re}}  - \alpha\, \Pi_{\mathrm{Re}} 
= \beta^i \partial_i \phi_{\mathrm{Re}} - \alpha\, \Pi_{\mathrm{Re}}
.\end{aligned}
\end{equation}

\subsection*{Evolution of $\Pi$}

We next derive the evolution equation for $\Pi_{\mathrm{Re}}$ in the $2$D Cartoon reduction. 
Starting from the standard $3+1$ form, we split spatial indices as $I=(i,a)$ with $i\in\{x,y\}$ 
on the evolution plane and $a=z$ the suppressed direction. On the $z=0$ plane, axisymmetry implies 
$\partial_a(\cdot)=0$ for scalar quantities such as $\phi_{\mathrm{Re}}$, $\alpha$, and $\chi$, so that 
terms containing $\partial_a\phi_{\mathrm{Re}}$, $\partial_a\alpha$, or $\partial_a\chi$ vanish, and mixed 
$(i,a)$ contributions drop out. However, second derivatives along the suppressed direction do not vanish 
identically and are replaced by the Cartoon identity $\partial_a\partial_a \phi_{\mathrm{Re}}
=\partial_z^2\phi_{\mathrm{Re}}=(\partial_y\phi_{\mathrm{Re}})/y$, which gives rise to the characteristic 
$1/y$ term in the final reduced equation.

\begin{equation}
{\setlength{\jot}{6pt}%
\begin{aligned}
\partial_t \Pi_{\mathrm{Re}}
=& \beta^I \partial_I \Pi_{\mathrm{Re}}
+ \alpha K \Pi_{\mathrm{Re}}
+ \alpha \chi \Gamma^I \partial_I \phi_{\mathrm{Re}}
+ \tilde{\gamma}^{IJ}
\left[
  -\chi (\partial_I \alpha)(\partial_J \phi_{\mathrm{Re}})
  + \alpha\left(
      \tfrac{1}{2} \partial_I \chi \partial_J \phi_{\mathrm{Re}} 
      - \chi\, \partial_I \partial_J \phi_{\mathrm{Re}}
    \right)
\right] + \alpha\, V^{\prime} \phi_{\mathrm{Re}} \\
=& \beta^i \partial_i \Pi_{\mathrm{Re}} + \beta^a \partial_a \Pi_{\mathrm{Re}} + \alpha K \Pi_{\mathrm{Re}} +  \alpha \chi \Gamma^i \partial_i \phi_{\mathrm{Re}}+\alpha \chi \Gamma^a \partial_a \phi_{\mathrm{Re}}+ 
\alpha V^{\prime} \phi_{\mathrm{Re}} \\
&-\chi \tilde{\gamma}^{ij}\partial_i\alpha\partial_j\phi_{\mathrm{Re}}-\chi \tilde{\gamma}^{ab}\partial_a\alpha\partial_b\phi_{\mathrm{Re}}-2\chi \tilde{\gamma}^{ib}\partial_i\alpha\partial_b\phi_{\mathrm{Re}}\\
&+\frac{\alpha}{2}\tilde{\gamma}^{ij}\partial_i\chi\partial_j\phi_{\mathrm{Re}}+\frac{\alpha}{2}\tilde{\gamma}^{ab}\partial_a\chi\partial_b\phi_{\mathrm{Re}}+ \alpha\tilde{\gamma}^{ib}\partial_i\chi\partial_b\phi_{\mathrm{Re}} \\
&-\alpha\chi\tilde{\gamma}^{ij}\partial_i\partial_j\phi_{\mathrm{Re}}-\alpha\chi\tilde{\gamma}^{ab}\partial_a\partial_b\phi_{\mathrm{Re}}-2\alpha\chi\tilde{\gamma}^{ib}\partial_i\partial_b\phi_{\mathrm{Re}} \\
&= \beta^i \partial_i \Pi_{\mathrm{Re}}
  - \chi \tilde{\gamma}^{ij}\partial_i\alpha\,\partial_j\phi_{\mathrm{Re}}
  + \alpha\Big[
      K \Pi_{\mathrm{Re}}
      + V^{\prime} \phi_{\mathrm{Re}}
      + \frac{1}{2}\tilde{\gamma}^{ij}\partial_i\chi\,\partial_j\phi_{\mathrm{Re}}
      + \chi\Big(
          \Gamma^i \partial_i \phi_{\mathrm{Re}}
          - \tilde{\gamma}^{ij}\partial_i\partial_j\phi_{\mathrm{Re}}
          - \tilde{\gamma}^{zz}\frac{\partial_y\phi_{\mathrm{Re}}}{y}
        \Big)
    \Big]
\end{aligned}}
\end{equation}
where $V^{\prime}=\frac{\partial V}{\partial |\phi|^{2}}$.

The expressions for $\partial_t \phi_{\mathrm{Im}}$ and $\partial_t \Pi_{\mathrm{Im}}$ follow analogously, with $\phi_{\mathrm{Re}}\rightarrow\phi_{\mathrm{Im}}$ and $\Pi_{\mathrm{Re}}\rightarrow\Pi_{\mathrm{Im}}$.

\subsection*{The expression for $\rho$}
We now derive the energy density $\rho=T_{\mu\nu}n^\mu n^\nu$ for the complex scalar field in the $2$D Cartoon reduction. It is convenient to introduce the following real, manifestly positive-definite combinations, written in terms of the evolved variables and spatial derivatives on the $z=0$ plane:
\begin{equation}
|\Pi|^{2}
=\Pi_{\mathrm{Re}}^{2} + \Pi_{\mathrm{Im}}^{2}.
\end{equation}

\begin{equation}
{\setlength{\jot}{6pt}%
\begin{aligned}
|\nabla\phi|^{2}
=& \chi\, \tilde{\gamma}^{IJ}
\left(
  \partial_I \phi_{\mathrm{Re}}\, \partial_J \phi_{\mathrm{Re}}
  + \partial_I \phi_{\mathrm{Im}}\, \partial_J \phi_{\mathrm{Im}}
\right) \\
=& \chi\, \tilde{\gamma}^{ij}
\left(
  \partial_i \phi_{\mathrm{Re}}\, \partial_j \phi_{\mathrm{Re}}
  + \partial_i \phi_{\mathrm{Im}}\, \partial_j \phi_{\mathrm{Im}}
\right) 
+2\chi\, \tilde{\gamma}^{ib}
\left(
  \partial_i \phi_{\mathrm{Re}}\, \partial_b \phi_{\mathrm{Re}}
  + \partial_i \phi_{\mathrm{Im}}\, \partial_b \phi_{\mathrm{Im}}
\right)
+\chi\, \tilde{\gamma}^{ab}
\left(
  \partial_a \phi_{\mathrm{Re}}\, \partial_b \phi_{\mathrm{Re}}
  + \partial_a \phi_{\mathrm{Im}}\, \partial_b \phi_{\mathrm{Im}}
\right) \\
=& \chi\, \tilde{\gamma}^{ij}
\left(
  \partial_i \phi_{\mathrm{Re}}\, \partial_j \phi_{\mathrm{Re}}
  + \partial_i \phi_{\mathrm{Im}}\, \partial_j \phi_{\mathrm{Im}}
\right) 
.\end{aligned}}
\end{equation}
On the $z=0$ plane, axisymmetry implies $\partial_a\phi_{\mathrm{Re}}=\partial_a\phi_{\mathrm{Im}}=0$ for scalar quantities, and therefore the mixed $(i,a)$ and purely $(a,b)$ contributions vanish. In this sense, the kinetic combinations $|\Pi|^{2}$ and $|\nabla\phi|^{2}$ retain the same algebraic form under dimensional reduction, provided one evaluates them using the reduced variables on the evolution plane.


\begin{equation}
\begin{aligned}
\rho
= \frac{1}{2} \bigl(|\Pi|^{2} + |\nabla\phi|^{2}\bigr)
 + \frac{1}{2} V(|\phi|^{2}) 
=\frac12\Bigl(
    \Pi_{\mathrm{Re}}^{2} + \Pi_{\mathrm{Im}}^{2}
  + \chi\, \tilde{\gamma}^{ij}
    \bigl(
        \partial_{i}\phi_{\mathrm{Re}}\,\partial_{j}\phi_{\mathrm{Re}}
      + \partial_{i}\phi_{\mathrm{Im}}\,\partial_{j}\phi_{\mathrm{Im}}
    \bigr)
  + V(|\phi|^{2})
  \Bigr)
\end{aligned}
\end{equation}

\subsection*{The expression for $S_I$}

We next consider the momentum density $S_I \equiv -\gamma^{\nu}{}_{I}T_{\mu\nu}n^{\mu}$. For a complex scalar field written in terms of its real and imaginary components, this projection takes a simple bilinear form involving the conjugate momentum $\Pi$ and the spatial gradients of the field. In the $2$D Cartoon reduction on the $z=0$ plane, we again split $I=(i,a)$ with $i\in\{x,y\}$ labeling directions on the evolution plane and $a=z$ denoting the suppressed direction. Since $\phi_{\mathrm{Re}}$ and $\phi_{\mathrm{Im}}$ are scalars, axisymmetry implies $\partial_a\phi_{\mathrm{Re}}=\partial_a\phi_{\mathrm{Im}}=0$ on $z=0$, and therefore $S_a=0$ so that only the in-plane components $S_i$ remain.

\begin{equation}
\begin{aligned}
S_I
=&~ \Pi_{\mathrm{Re}}\, \partial_I \phi_{\mathrm{Re}}
 + \Pi_{\mathrm{Im}}\, \partial_I \phi_{\mathrm{Im}} \,.
\end{aligned}
\end{equation}
After splitting $I=(i,a)$, the components are
\begin{equation}
\begin{aligned}
S_i
=&~ \Pi_{\mathrm{Re}}\, \partial_i \phi_{\mathrm{Re}}
 + \Pi_{\mathrm{Im}}\, \partial_i \phi_{\mathrm{Im}} \,,\\
S_a
=&~ \Pi_{\mathrm{Re}}\, \partial_a \phi_{\mathrm{Re}}
 + \Pi_{\mathrm{Im}}\, \partial_a \phi_{\mathrm{Im}}
 = 0\,.
\end{aligned}
\end{equation}

\subsection*{The expression for $S_{IJ}$}

We next derive the spatial stress tensor $S_{IJ}\equiv \gamma^{\mu}{}_{I}\gamma^{\nu}{}_{J}T_{\mu\nu}$ for the complex scalar field. Written in terms of the real and imaginary parts, $S_{IJ}$ contains a direction-dependent gradient contribution, $\partial_I\phi\,\partial_J\phi$, together with an isotropic term proportional to the spatial metric $\tilde{\gamma}_{IJ}$. In the $2$D Cartoon reduction on the $z=0$ plane, we split $I=(i,a)$, where $i\in\{x,y\}$ labels directions on the evolution plane and $a$ denotes the suppressed direction. For scalar quantities, axisymmetry implies $\partial_a\phi_{\mathrm{Re}}=\partial_a\phi_{\mathrm{Im}}=0$ on $z=0$, so the gradient contribution to the mixed components $S_{ia}$ vanishes. Moreover, in our axisymmetric (nonspinning) setup the mixed conformal-metric components satisfy $\tilde{\gamma}_{ia}=0$ on the evolution plane, so the metric-proportional part does not contribute either and therefore $S_{ia}=S_{ai}=0$. The nontrivial components are thus the in-plane stresses $S_{ij}$. For the suppressed block $S_{ab}$, the gradient term $\partial_a\phi\,\partial_b\phi$ also vanishes on $z=0$, leaving $S_{ab}$ determined entirely by the term proportional to $\tilde{\gamma}_{ab}$.
\begin{equation}
\begin{aligned}
S_{I J}
=& \partial_I \phi_{\mathrm{Re}} \partial_J \phi_{\mathrm{Re}}
+ \partial_I \phi_{\mathrm{Im}} \partial_J \phi_{\mathrm{Im}}
- \frac{1}{2\chi}\,\tilde{\gamma}_{I J}\bigl(|\nabla\phi|^{2} - |\Pi|^{2} + V(|\phi|^{2})\bigr)\,,
\end{aligned}
\end{equation}
and after splitting $I=(i,a)$ one obtains, for the different blocks,
\begin{equation}
{\setlength{\jot}{6pt}
\begin{aligned}
S_{ij}
=& \partial_i \phi_{\mathrm{Re}} \partial_j \phi_{\mathrm{Re}}
+ \partial_i \phi_{\mathrm{Im}} \partial_j \phi_{\mathrm{Im}}
- \frac{1}{2\chi}\,\tilde{\gamma}_{i j}\bigl(|\nabla\phi|^{2} - |\Pi|^{2} + V(|\phi|^{2})\bigr)\,,\\
S_{ia}
=&~0\,,\\
S_{ab}
=& - \frac{1}{2\chi}\,\tilde{\gamma}_{a b}\bigl(|\nabla\phi|^{2} - |\Pi|^{2} + V(|\phi|^{2})\bigr)\,
\end{aligned}}
\end{equation}
where the gradient term in $S_{ab}$ vanishes on the evolution plane because $\partial_a\phi_{\mathrm{Re}}=\partial_a\phi_{\mathrm{Im}}=0$ for scalars. In our setup there is only one suppressed direction, so one may finally identify $a=b=z$ and obtain $S_{ab}\to S_{zz}$.

\subsection*{The expression for $S$.}
We now obtain the scalar quantity $S\equiv \gamma^{IJ}S_{IJ}$, i.e., the trace of the spatial stress tensor on the hypersurface. In the conformal BSSN/CCZ4 variables used here, this trace can be written as $S=\chi\,\tilde{\gamma}^{IJ}S_{IJ}$. In the $2$D Cartoon reduction we split indices as $I=(i,a)$ with $i\in\{x,y\}$ and $a=z$, so the contraction separates into in-plane and suppressed contributions. The final factor of $3/2$ arises purely from this trace operation: on the spatial slice there are three directions, and therefore $\tilde{\gamma}^{ij}\tilde{\gamma}_{ij}=2$ while $\tilde{\gamma}^{zz}\tilde{\gamma}_{zz}=1$, giving $\frac{1}{2}\bigl(\tilde{\gamma}^{ij}\tilde{\gamma}_{ij}+\tilde{\gamma}^{zz}\tilde{\gamma}_{zz}\bigr)=\frac{3}{2}$.

\begin{equation}
{\setlength{\jot}{6pt}%
\begin{aligned}
S
=& \chi\ \tilde{\gamma}^{I J} S_{I J} \\
=& \chi\ \tilde{\gamma}^{i j} S_{i j} + \chi\ \tilde{\gamma}^{a b} S_{a b} \\
=& \chi\ \tilde{\gamma}^{i j} \partial_i \phi_{\mathrm{Re}} \partial_j \phi_{\mathrm{Re}} + \chi\ \tilde{\gamma}^{i j} \partial_i \phi_{\mathrm{Im}} \partial_j \phi_{\mathrm{Im}}
+\frac{1}{2}\Bigl(\tilde{\gamma}^{i j}\tilde{\gamma}_{i j}  + \tilde{\gamma}^{a b}\tilde{\gamma}_{a b}\Bigr)
\Bigl(-V(|\phi|^{2}) - |\nabla\phi|^{2} + |\Pi|^{2}\Bigr) \\
=& \chi\ \tilde{\gamma}^{i j} \partial_i \phi_{\mathrm{Re}} \partial_j \phi_{\mathrm{Re}} + \chi\ \tilde{\gamma}^{i j} \partial_i \phi_{\mathrm{Im}} \partial_j \phi_{\mathrm{Im}}
+\frac{1}{2}\Bigl(\tilde{\gamma}^{i j}\tilde{\gamma}_{i j}  + \tilde{\gamma}^{z z}\tilde{\gamma}_{z z}\Bigr)
\Bigl(-V(|\phi|^{2}) - |\nabla\phi|^{2} + |\Pi|^{2}\Bigr) \\
=& \chi\,\tilde{\gamma}^{ij}\bigl(\partial_i \phi_{\mathrm{Re}} \partial_j \phi_{\mathrm{Re}} + \partial_i \phi_{\mathrm{Im}} \partial_j \phi_{\mathrm{Im}}\bigr) + \frac{3}{2}\bigl(|\Pi|^{2} - |\nabla\phi|^{2} - V(|\phi|^{2})\bigr)
\end{aligned}}
\end{equation}


\section{Constraint and Convergence}
\label{app:Cons}

The key results reported in this paper are obtained from numerical-relativity simulations, in particular the gravitational-wave energy extracted from the dynamical spacetime. It is therefore important to demonstrate that (i) the evolution maintains acceptable levels of constraint violation and (ii) the extracted radiated energy is robust against changes in numerical resolution. In this appendix we provide representative diagnostics for the configuration with $\lambda=10$ and $|\phi_c|=0.02$.

\begin{figure}
\centering
\includegraphics[width=0.4\textwidth]{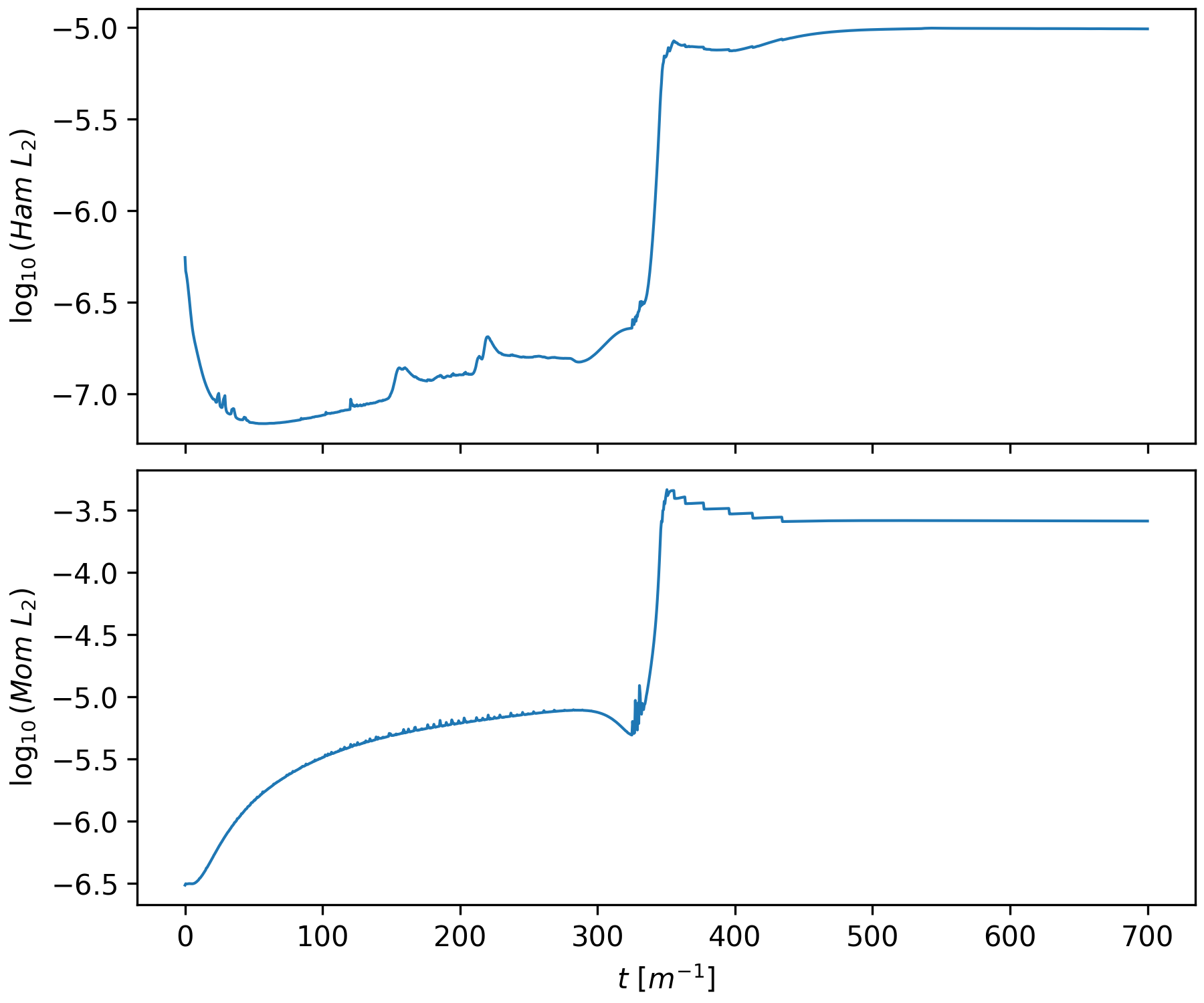}
\caption{Constraint violation for $\lambda=10$ and $|\phi_c|=0.02$ at resolution $N=256$. The upper panel shows the Hamiltonian constraint and the lower panel shows the momentum constraint. The vertical axis is $\log_{10}$ of the corresponding constraint norm, and the horizontal axis is the simulation time. The binary coalesces at approximately $t\simeq 300$.}\label{fig:Constraint}
\end{figure}

Fig.~\ref{fig:Constraint} shows the time evolution of the Hamiltonian and momentum constraint norms at resolution $N=256$. A nonzero constraint level is expected already at early times because the binary initial data are constructed by superposition rather than by directly solving the full constraint equations. During the inspiral the constraints remain bounded, while the merger phase (occurring here at approximately $t\simeq 300$) is accompanied by a transient increase associated with the development of strong gradients and gauge dynamics.

\begin{figure}
\centering
\includegraphics[width=0.4\textwidth]{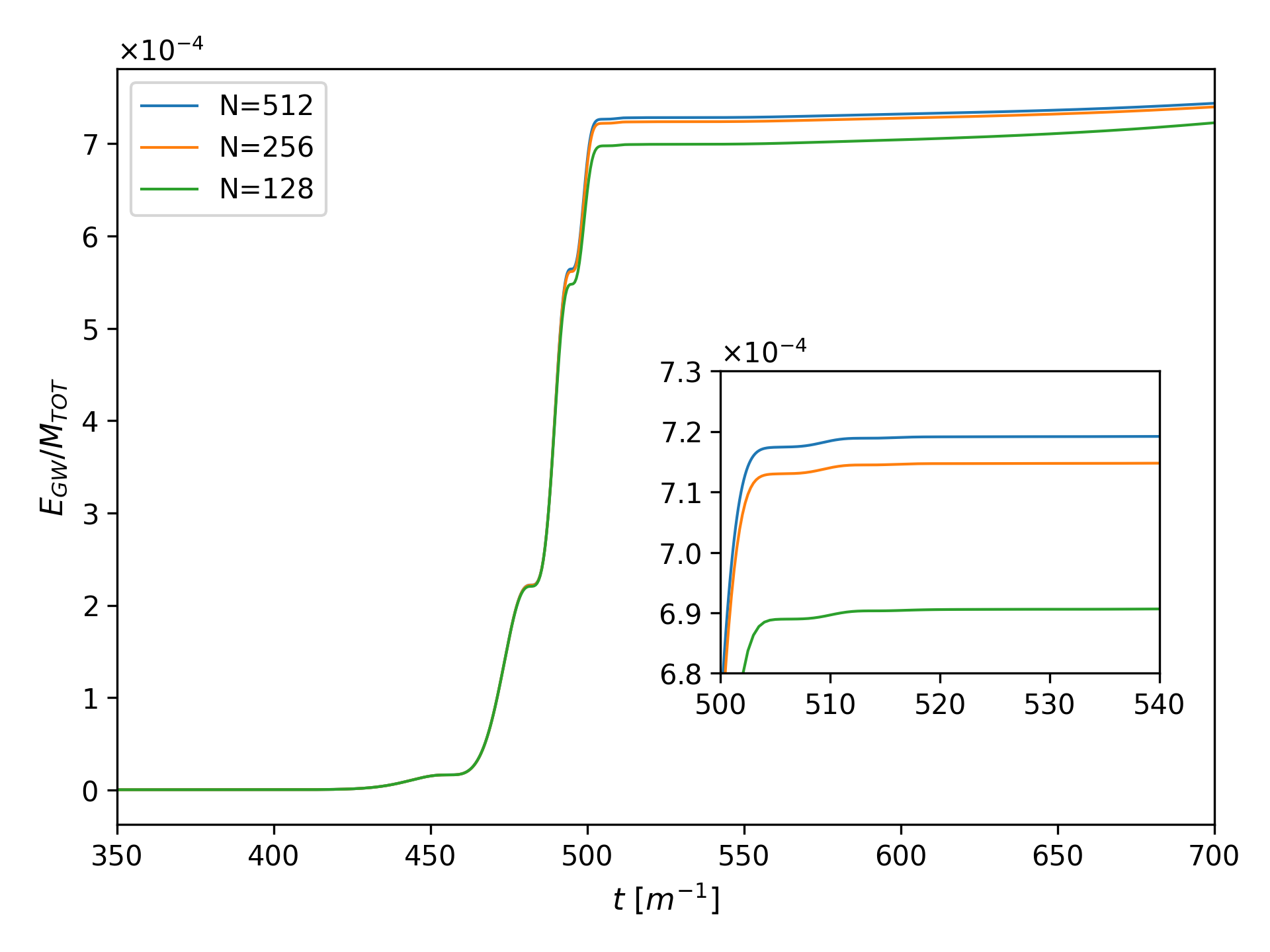}
\caption{Convergence of the gravitational-wave energy for $\lambda=10$ and $|\phi_c|=0.02$ at different resolutions $N$, where $N$ denotes the number of grid points in each spatial direction. The radiated energy on the vertical axis is normalized by the initial total mass $M_{\rm TOT}$.}\label{fig:EnergyConv}
\end{figure}

To assess the impact of finite resolution on the gravitational-wave energetics, Fig.~\ref{fig:EnergyConv} compares the normalized radiated energy $E_{\rm GW}/M_{\rm TOT}$ obtained at several resolutions. The overall behaviour and final radiated energy are consistent across resolutions, and the residual differences between successive resolutions provide a practical estimate of the discretization uncertainty for the representative case shown here.


\clearpage
\bibliography{BS}

\end{document}